\pdfoutput=1

\documentclass[final,5p,times,twocolumn,english]{elsarticle}
\usepackage[T1]{fontenc}
\usepackage[latin9]{inputenc}
\usepackage{multirow}
\usepackage[english]{babel}
\usepackage{graphicx}
\usepackage{verbatim}
\usepackage[format=plain,justification=centerlast,width=.8\textwidth]{caption}
\usepackage{subcaption}
\usepackage{mathrsfs}
\usepackage{bbold}
\usepackage{amsmath}
\usepackage{xcolor}
\usepackage{xspace}
\usepackage{dcolumn} % align table columns on decimal point
\usepackage{color,soul}
\usepackage[colorlinks,linkcolor=blue,citecolor=blue,urlcolor=blue,breaklinks=true,draft]{hyperref}

\newcommand{\geant}{\textsc{Geant4}\xspace}
\newcommand{\FIG}{Figure~}
\newcommand{\FIGS}{Figures~}
\newcommand{\WT}{\xspace wt\%\xspace}
\newcommand{\setupthecaption}{\captionsetup{width=0.95\columnwidth}}
\def \myfigurewidth{1}
\newcolumntype{p}[1]{D{,}{\,\pm\,}{#1}} % aligns plus/minus (ex 13.5,0.1)
\newcolumntype{d}[1]{D{,}{.}{#1}} % align decimal point (ex 1,09)

\newcommand{\CITE}[1]{\protect{\cite{#1}}}

\begin{document}
\begin{frontmatter}

%%BEGIN DOCUMENT %%
\title{Characterization of Gadolinium-loaded Plastic Scintillator for Use as a Neutron Veto}

%\input{author_list}
%\author{D.M. Poehlmann, D. Barker, H. Chagani, P. Cushman, G. Heuermann, A. Medved, H.~E. Rogers, R. Schmitz}

	%\email{degui008@umn.edu} %for revtex
	 %for elsarticle
\author[add1]{D.M. Poehlmann\corref{cor1}}
\ead{degui008@umn.edu}
\author[add1]{D. Barker}
\author[add2]{H. Chagani}
\author[add1]{P. Cushman}
\author[add3]{G. Heuermann}
\author[add1]{A. Medved}
\author[add1]{H.~E. Rogers}
\author[add1]{R. Schmitz}
\date{\today}

\cortext[cor1]{Corresponding author}
\address[add1]{School of Physics and Astronomy, University of Minnesota, Minneapolis, MN 55455, USA}
\address[add2]{Diamond Light Source, Didcot, United Kingdom}
\address[add3]{Karlsruhe Institute of Technology, Karlsruhe, Germany}

\begin{abstract}
Scintillator doped with a high neutron-capture cross-section material can be used to detect neutrons via their resulting gamma rays. Examples of such detectors using liquid scintillator have been successfully used in high-energy physics experiments. However, a liquid scintillator can leak and is not as amenable to modular or complex shapes as a solid scintillator. Polystyrene-based scintillators from a variety of gadolinium compounds  with varying concentrations were polymerized in our laboratory.  The light output, emission spectra, and attenuation length of our samples were measured and light collection strategies using a wavelength shifting (WLS) fiber were evaluated.  The measured optical parameters were used to tune a \geant-based optical Monte Carlo, enabling the trapping efficiency to be calculated.  This technology was also evaluated as a possible neutron veto for the direct detection dark matter experiment, Super Cryogenic Dark Matter Search (SuperCDMS). 
\end{abstract}

\begin{keyword}
CDMS \sep dark matter \sep high energy physics \sep neutron \sep plastic scintillator \sep veto 
\end{keyword}

%\maketitle

\end{frontmatter}

% path for all DMP macros: /data/zlatibor/cdms/poehlmann/scintillator_characterization_paper/

%%BEGIN SECTIONS
\section{Introduction}

As dark matter experiments explore ever-larger exposures in their search for a direct detection of weakly-interacting massive particles (WIMPs), background rejection becomes increasingly difficult. Strategies to reject electron recoils have been effective in reducing gamma and beta backgrounds by orders of magnitude, but neutron identification remains difficult.  Since neutrons, like WIMPs, interact via nuclear recoil, only those neutrons that produce interactions in multiple detectors due to their much shorter mean free path can be reliably eliminated.  Neutron-produced single-scatter nuclear recoils remain an indistinguishable background.  Most underground experiments thus rely on eliminating neutron background events through shielding. Since a large sample of neutrons cannot be identified in situ, experiments rely on extensive simulations to quantify the reduction achieved.   An inexpensive and efficient neutron veto would be useful in directly rejecting incident neutrons, as well as understanding the shield efficiency and identifying internal contaminants. 

Neutron detectors based on liquid scintillator have been effective in a number of experiments \CITE{SuperK,darkside,SNO,KamLAND,Borexino}, but for experiments with a more complicated geometry, liquid scintillator can be problematic, since forming multiple small containment vessels can be expensive and the issue of leakage is always present. A solid neutron veto is easily machined, installed, and maintained.   Furthermore, the dopants frequently used in liquid scintillators, such as trimethyl borate (TMB), are dangerous and highly flammable \CITE{TMB_SDS}, but can be stabilized when polymerized in a solid.

The SuperCDMS experiment~\CITE{CDMS_SNOLAB_sensitivities} will be installed at SNOLAB in Sudbury Canada.  The experiment will measure both ionization and phonon signals from crystals of germanium and silicon cooled to tens of mK. The cryostat will be connected to a helium dilution refrigerator, surrounded by shielding composed of layers of lead and high-density polyethylene.  The initial payload will consist of four towers of detectors, for which the neutron background will be negligible.   For larger payloads, however, there will be neutron-induced events that can mimic WIMPs.   As an upgrade, the innermost layer of polyethylene could be replaced by an active layer of gadolinium-doped plastic scintillator embedded with wavelength-shifting (WLS) fibers and read out using radio-pure silicon photomultipliers (SiPMs). The work reported here was part of a feasibility study for such an upgrade.

\section{Gd-loaded Scintillator Sample Production}

\subsection{Organic Scintillator as a Neutron Detector}
\begin{figure}[t]
	\centering
	\setupthecaption
	\includegraphics[width = \myfigurewidth \columnwidth]{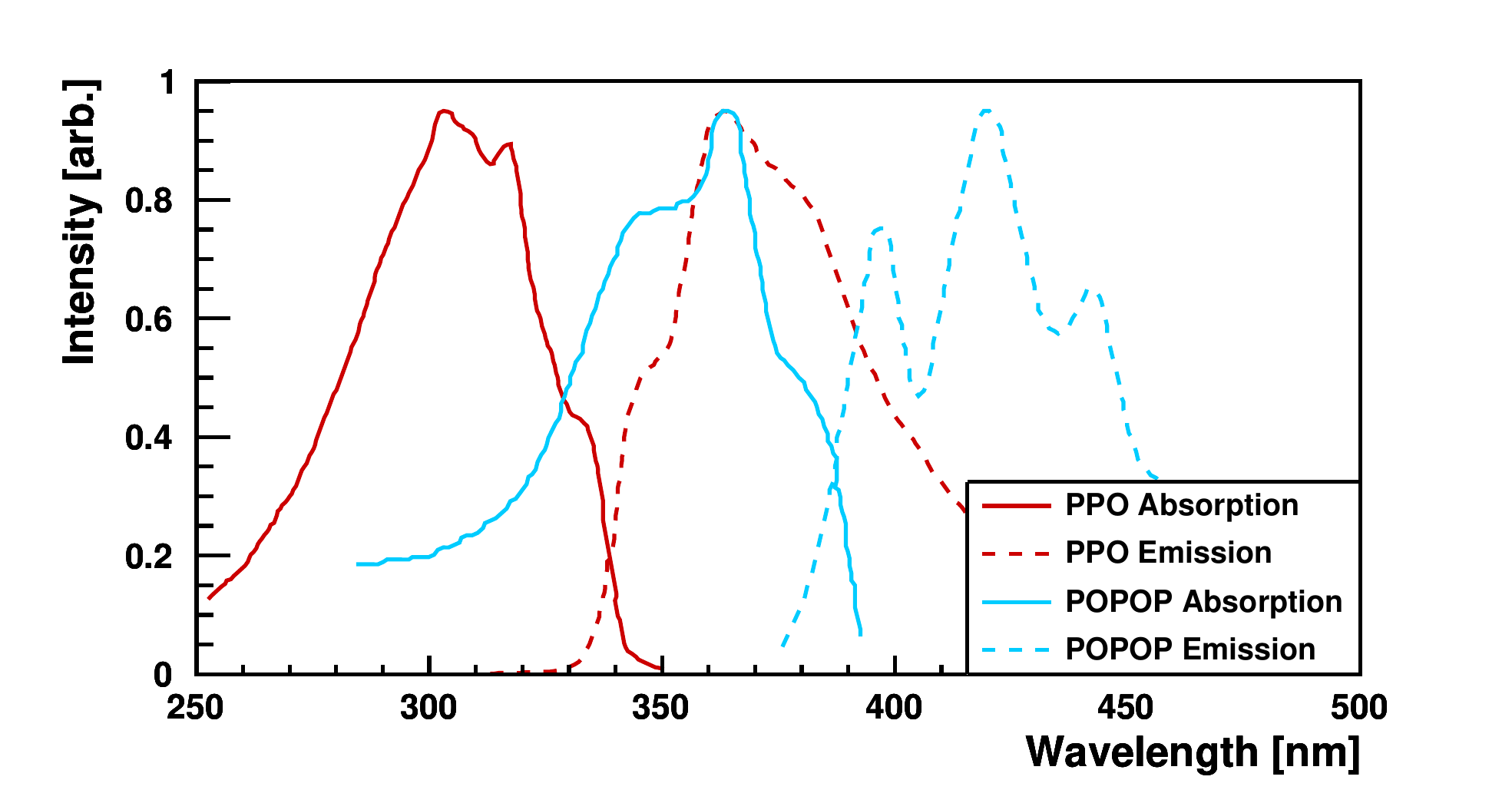}
	\caption{The absorption and emission spectra of the fluorescent dyes used in the production of the scintillators. Figure adapted from~\CITE{Berlman}.}
	\label{fig:fluorSpectra}
\end{figure}

Solid organic scintillators are made by polymerizing monomers such as styrene or vinyl toluene.  There are many types of fluorescent dyes that can be added during this process.  The primary fluorescent dye compound used here is 2,5-diphenyloxazole (PPO), and the secondary dye is 1,4-bis(5-phenyloxazol-2-yl) benzene (POPOP).   The PPO acts as a pump for the POPOP, which fluoresces naturally with a maximum intensity at 429~nm~\CITE{Breukers} (see \FIG \ref{fig:fluorSpectra}), well-matched to the quantum efficiency of commonly available photomultiplier tubes.   

\begin{figure}[t]
	\centering
	\setupthecaption
	\includegraphics[width = \myfigurewidth \columnwidth]{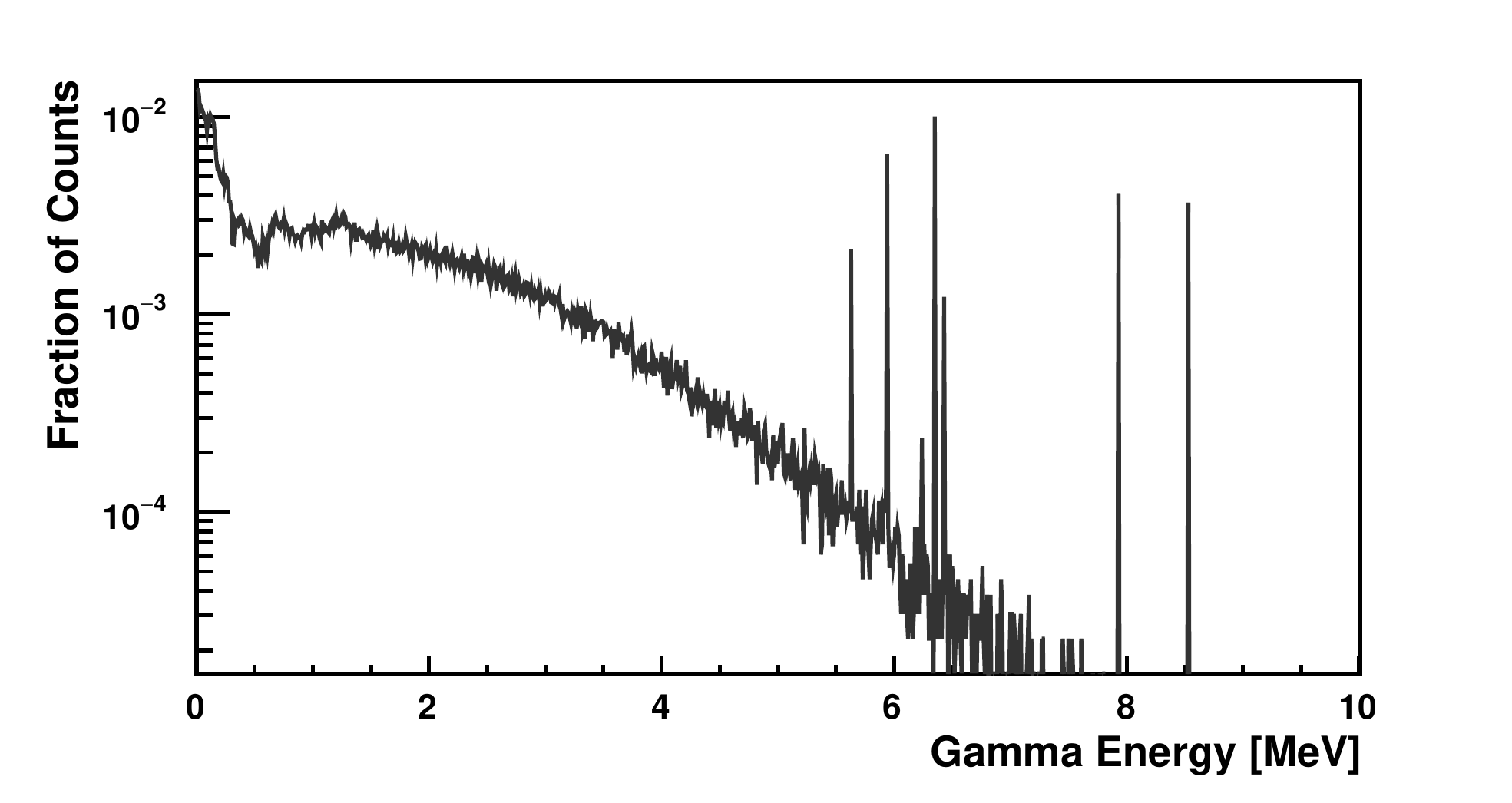} 
	\caption{$\gamma$-ray energy spectrum from neutron capture on gadolinium. }
	\label{fig:gd_gammas}
\end{figure}

Additional doping of the scintillator with a material of high neutron-capture cross-section allows for neutron detection via the resulting secondary particles. Compounds containing gadolinium, lithium, and boron are commonly used. Two gadolinium (Gd) isotopes have the highest neutron absorption cross section of any naturally occurring element: $^{155}$Gd (14.7\% abundance) with a cross section of $6.1\times10^{-20}$~cm$^2$ and $^{157}$Gd (15.7\% abundance) with a cross-section of $2.6\times10^{-19}$~cm$^2$~\CITE{pdg}.  Interactions between neutrons and gadolinium result in low-energy internal conversion electrons and a cascade of Auger electrons and gamma-rays.  Atomic spectral x-rays are also emitted 50\% of the time. $^{157}$Gd emits a cascade of gammas (on average three) with a total energy of 7.9~MeV, while $^{155}$Gd releases a total of 8.5~MeV energy in the capture process~\CITE{kandlakunta2013measurement}. \FIG \ref{fig:gd_gammas} shows a spectrum of individual gamma energies generated by neutrons captured on Gd, weighted by relative isotope abundance.  The spectrum was produced by the \geant Monte Carlo simulation described in Section~\ref{section:Monte_Carlo}.


\subsection{Production Methods}

The styrene polymerization process described below roughly follows the method described by Bell \textit{et al.\ }\CITE{Belletal}. Prior to polymerization, the internal inhibitor (added to the styrene monomer by the manufacturer to improve shelf life)  was removed using a column of alumina. The styrene monomer was then combined with 1\WT PPO and 0.1\WT POPOP.   Purified benzoyl peroxide (added in at 0.8\WT) was used as a catalyst for polymerization. At this point, the metallic compound was added to the purified styrene in the desired weight percentage.  The resulting mixture was sonicated for up to an hour to encourage complete dissolution.  

If water is not removed from the system completely, bubbles form in the polystyrene during polymerization. To prevent this, dry nitrogen was bubbled through the solution for at least an hour to remove water and oxygen.  The styrene was polymerized in a $\sim$60$^{\circ}$C water bath for 5-6 days under nitrogen and slowly cooled to room temperature over 12 hours. The time for each step was optimized through multiple trials. Sample diameters were typically 2.5~cm, while individual sample lengths ranged from 2 - 11~cm. 

\begin{figure}
\centering
\setupthecaption
	\includegraphics[width = \myfigurewidth \columnwidth]{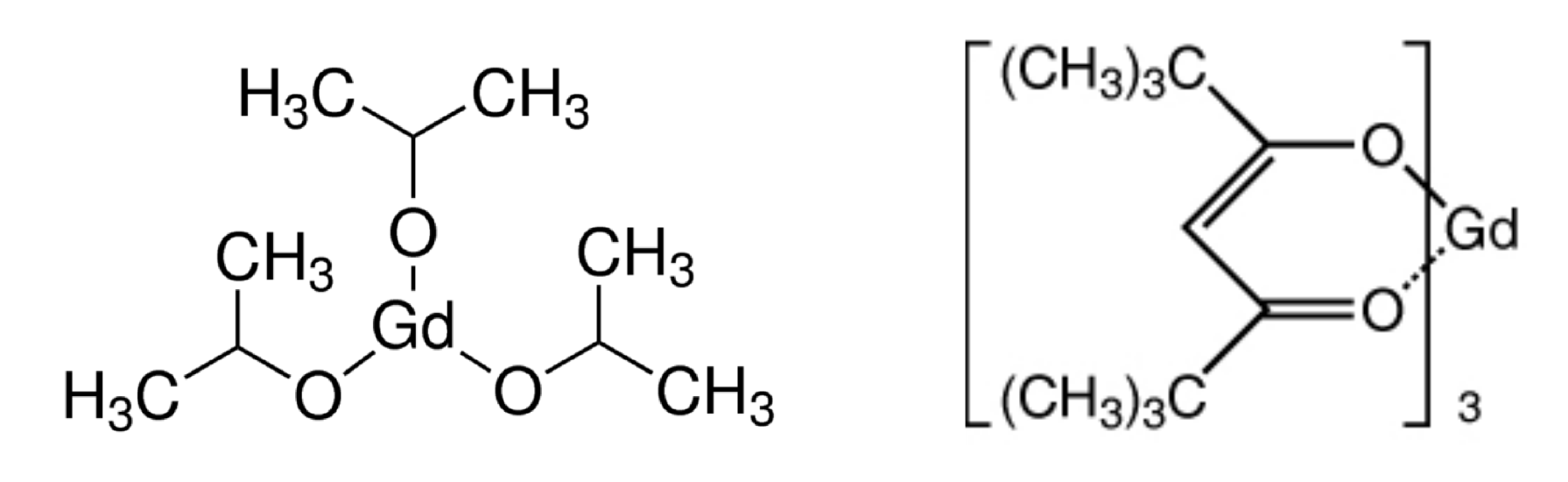}
	\caption{The chemical structure of gadolinium compounds used in our optical measurements. 
Left: Gd(i-Pr)$_3$;    Right: Gd(TMHD)$_3$}
\label{fig:gdChemStruct}
\end{figure}

Table~\ref{tab:gd_loading} summarizes the properties of some Gd-containing compounds that we polymerized in our lab. The best samples were obtained for gadolinium isopropoxide (Gd(i-Pr)$_3$)~\CITE{gdiso} and gadolinium 2,2,6,6-tetramethyl-3,5-heptanedione (Gd(TMHD)$_3$)~\CITE{gdtmhd}, whose chemical structures are both shown in \FIG \ref{fig:gdChemStruct}.

\begin{table}[t]
\centering
\setupthecaption
\resizebox{\columnwidth}{!}{%
\begin{tabular}{c|c|c|c}
Dopant & \WT Gd  & Cost & Notes \\ \hline
Gd(i-Pr)$_3$~\CITE{Ovechkina} & 47.01\% & \$\$\$ & Optical measurements \\ %sigma aldrich
Gd(TMHD)$_3$~\CITE{Bertrand} & 22.24\% & \$\$ & Optical measurements \\ %alfa aesar
GdB$_6$ & 70.80\% & \$\$ & Purple precipitate\\ %sigma aldrich
Gd(OTf)$_3$ & 26.01\% & \$ & Orange samples\\ %gd triflate
Gd(NO$_3$)$_3$$\cdot$6H$_2$O~\CITE{brudanin2001element} & 34.84\% &  \$ & Yellow samples
\end{tabular}%
}
\caption{Gd compounds used to dope scintillators polymerized in our laboratory.   If the compound was polymerized successfully elsewhere, we CITE the referenced paper. The percent mass of Gd per unit of compound is listed in the "wt\% Gd" column; the relative cost of each compound is based on\WT Gd.  A more extensive review of Gd-containing compounds for use as dopants in plastic scintillator can be found in J. Dumazert \textit{et al.\ }\CITE{dumazert2017gadolinium}.} 
\label{tab:gd_loading}
\end{table}

\subsection{Gadolinium Compound Trials}

 The samples used in the optical tests described below were made using the same batch of pre-made scintillator fluor cocktail plus catalyst.  All samples were 11.10$\pm0.02$~cm in length and 2.50$\pm0.02$~cm in diameter. Both the front and rear faces of these samples were cut and polished.  Maximum loadings of 0.071\WT Gd using Gd(i-Pr)$_3$ and 0.055\WT Gd using Gd(TMHD)$_3$ were achieved, which are both significantly less than expected from previous publications~\CITE{Ovechkina, Bertrand}.   Additional heating and sonication prior to the polymerization process did not improve the solubility of either compound, but instead decreased the quality of the resulting scintillator. While not explicitly mentioned, Bertrand \textit{et al.\ }\CITE{Bertrand} added unspecified polar co-monomers at 20-25\WT of the total monomers. It was hypothesized by Bertrand \textit{et al.\ }\CITE{BertrandPrivate} that partial ligand exchange or completion of the Gd coordination sphere with the polar co-monomers allowed for higher concentrations to be reached. However, the specific co-monomer used for Gd(TMHD)$_3$ is considered to be a trade secret. Ovechkina \textit{et al.\ }\CITE{Ovechkina} also did not provide complete information on their polymerization process in their published paper.  
 
In early trials with Gd(i-Pr)$_3$, cloudy samples were observed. This cloudiness was also reported by Ovechkina \textit{et al.\ }\CITE{Ovechkina}. Due to impurities in the metallic compound, a white, wispy precipitate was filtered out of the styrene-metal dopant solution. The mass of precipitate recovered was smaller than the uncertainty in the balance used to measure the doping compound, indicating that any Gd lost had a negligible effect on the calculated Gd concentration. Elemental analysis of the recovered precipitate using energy dispersive spectroscopy on a scanning electron microscope \footnote{Characterization and analysis on the (FEG-SEM)~JEOL 6700 by N. Seaton, http://www.charfac.umn.edu/instruments/, Shepherd Lab,  University of Minnesota.} showed its composition was inconsistent with that of Gd(i-Pr)$_3$, but was instead an oxide, most likely Gd$_2$O$_3$. Its presence may be due to impurities in the dopant or exposure to trace amounts of moisture.

After one hour of sonication in solution with styrene, PPO, POPOP, and BPO, the samples doped with Gd(TMHD)$_3$ fully dissolved at concentrations up to 0.75\WT Gd.  However, later the dopant precipitated out of the scintillator mixture during the polymerization process. For both compounds, only concentrations $<$~0.055\WT Gd reliably produced a final clear polymerized scintillator. Optical measurements were made on our samples for those concentrations where the dopant was fully dissolved, and are presented in the next section.  

\begin{figure}[t]
\centering
\setupthecaption
\begin{subfigure}{0.49\columnwidth}
	\includegraphics[width = \columnwidth]{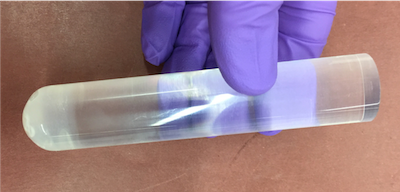}
	\caption{Gd(i-Pr)$_3$ (0.1\WT Gd)}
	\end{subfigure}
\begin{subfigure}{0.49\columnwidth}
	\includegraphics[width = \columnwidth]{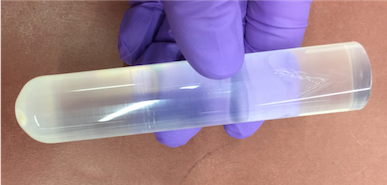}
	\caption{Gd(TMHD)$_3$ (0.1\WT  Gd)}
	\end{subfigure}
\caption{Overloaded Gd samples with reproducible precipitate layers.}
\label{fig:fab_samples}
\end{figure}

Overloaded samples with undissolved precipitate were also studied. Since the precipitate settles to the bottom, stacking non-uniform Gd-loaded scintillator panels would be similar to neutron detector geometries where layers of opaque Gd-loaded materials are interleaved with optically clear scintillator readout bars.  As shown in Section~\ref{section:veto_efficiency}, uniform doping is preferable, but reasonable neutron veto efficiencies can be achieved if the interleaving is fine enough.  Examples of overloaded samples are shown in \FIG \ref{fig:fab_samples}. The height of the precipitate is roughly linear with\WT doping within a narrow range of concentrations between precipitation and the point where the undissolved compound prevents the samples from fully polymerizing. 

\section{Light Output Measurements}

\subsection{ Characterizing the Photodetector}
The scintillation light was read out by a 2-inch Hamamatsu R1332 photomultiplier tube (PMT) with a green-sensitive photocathode. The PMT quantum efficiency, QE($\lambda$) was directly measured in order to construct a reliable Monte Carlo simulation of our samples' optical properties.   The QE was determined using
\begin{equation} \label{eqn:PMTqe}
\text{QE}(\lambda) = \frac{h c}{e} \frac{S(\lambda)}{\lambda},
\end{equation}
where $\frac{hc}{e}$ is 1240~V$\cdot$nm, $\lambda$ is the wavelength in nm, and $S(\lambda)$ is the radiant sensitivity of the PMT as a function of wavelength in A$\cdot$W$^{-1}$~\CITE{PMThandbook}. 
The PMT gain was measured at bias voltages between 1600~V and 2100~V by self-triggering at very low light levels and identifying the position of the single photoelectron peak. Over this voltage range, the gain curve was fit to
\begin{equation} \label{eqn:gain_fit_eqn}
G(V_b) = (6.29\pm0.13)e^{(4.782\pm0.073 \text{V}^{-1})V_b}  
\end{equation}
where $V_b$ is the PMT bias in kV, with a reduced $\chi^2$ of 0.894.

\begin{figure*}[t]
\centering
	\includegraphics[width = .7\textwidth]{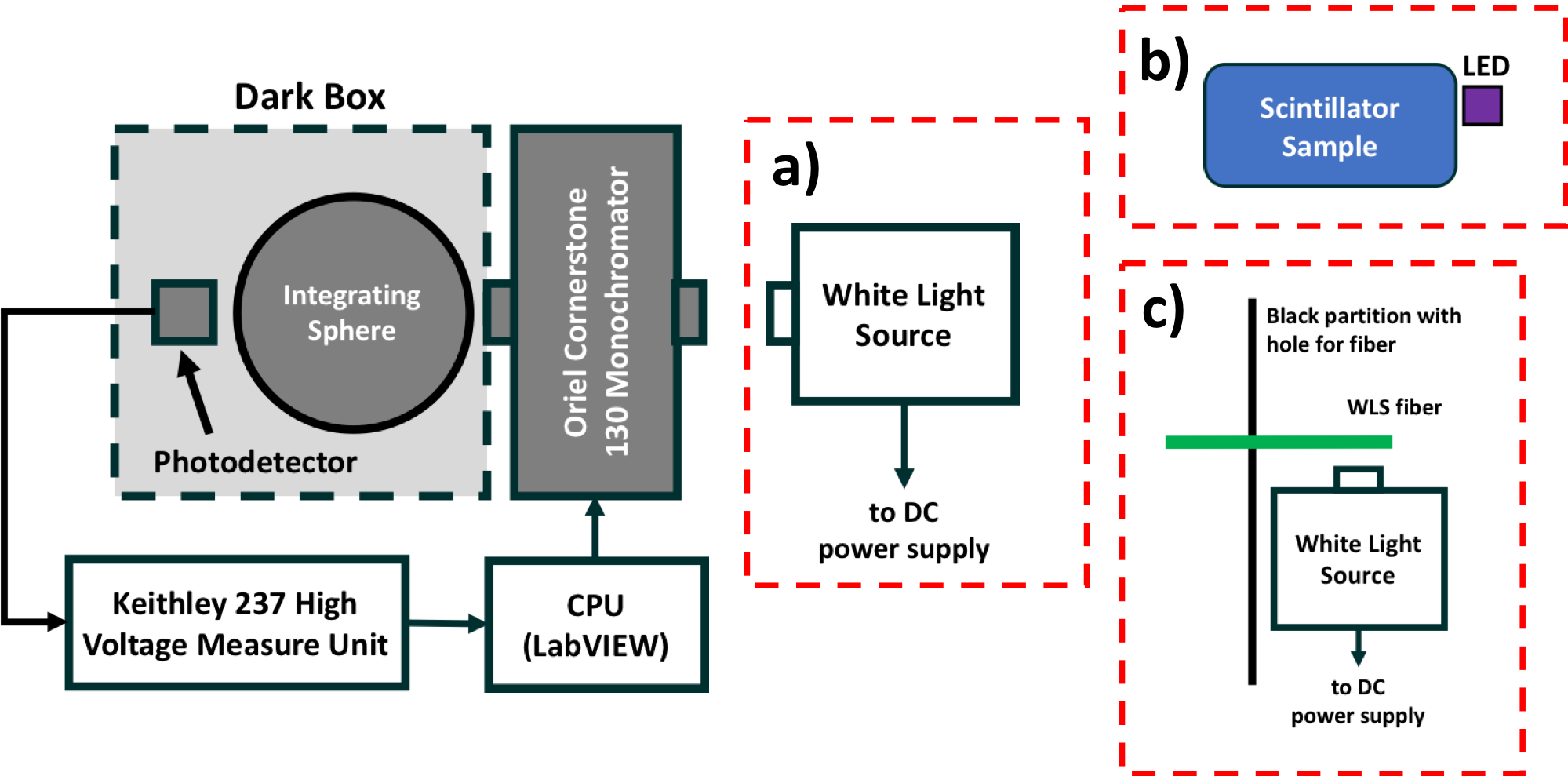}   
	\caption{Monochromator measurement setup with three possible input configurations. Configuration (a) was used to determine the quantum efficiency of the Hamamatsu R1332 PMT and the SiPM, as well as calibrating the white light source with a photodiode.  Configuration (b) and (c) were used to measure emission spectra of the blue scintillator and WLS fiber, respectively, with the calibrated photodiode as the photodetector.}
	\label{fig:MonoNew}
\end{figure*}

\begin{figure}[t]
    \centering
    \setupthecaption
    \includegraphics[width = \myfigurewidth \columnwidth]{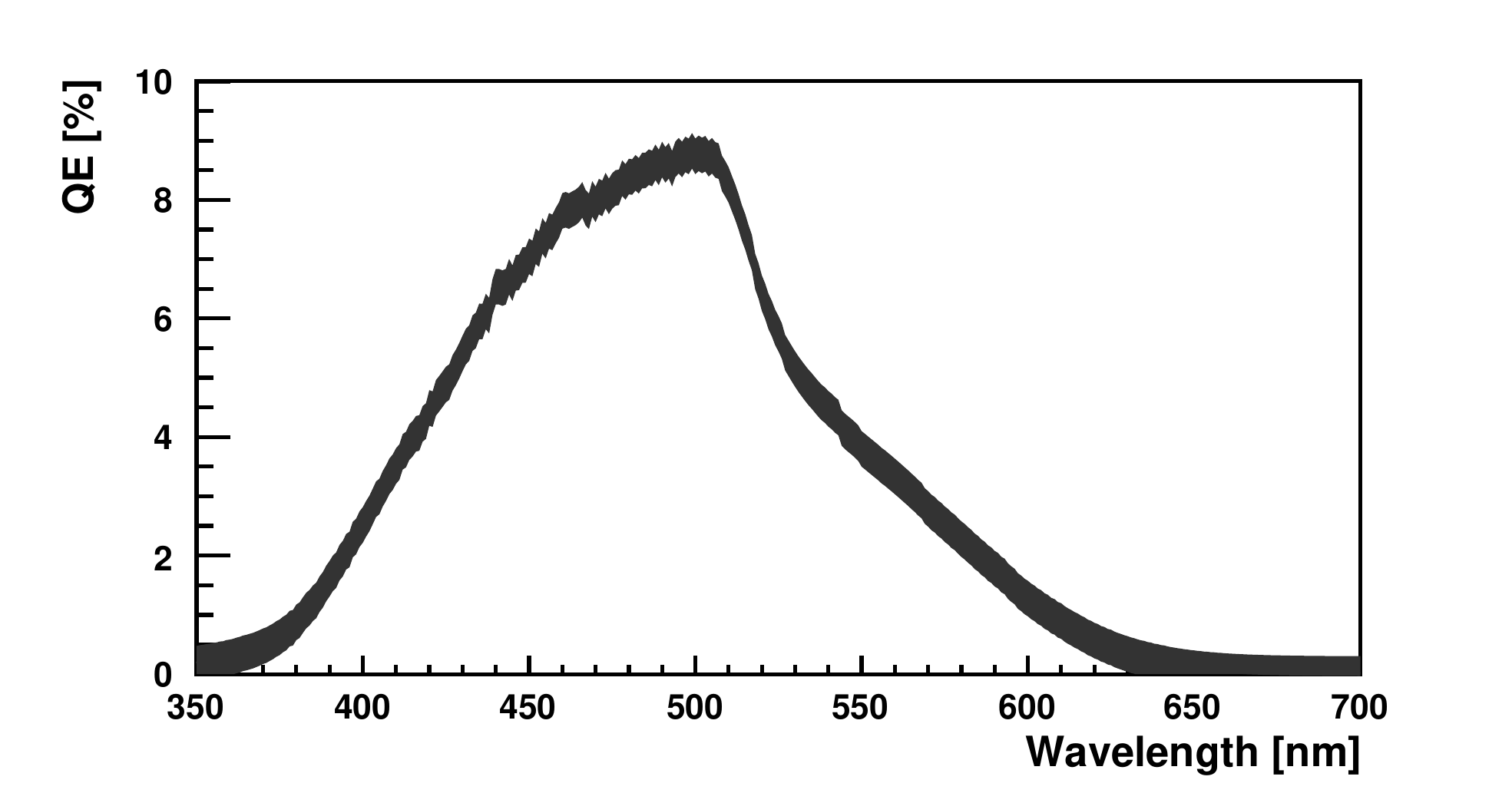} 
    \caption{Quantum efficiency of the PMT measured in monochromator Configuration (a). The line width is representative of the systematic uncertainties derived from the standard deviation of repeated trials.}
	\label{fig:mono_results}
\end{figure}

An Oriel Cornerstone 130 monochromator (see \FIG\ref{fig:MonoNew}) was used for all spectral measurements.   Configuration (a) was used to measure the radiant sensitivity of the PMT.  An integrating sphere was used to uniformly distribute light over the active surface of the photodetector. The spectrum of the white light source was then measured using both a calibrated photodiode (Newport 818-ST2-UV-DB) and the PMT biased to negative 1800 V.  The radiant sensitivity ($S$) at a given wavelength was then calculated using
\begin{equation} \label{eqn:PMTresponsivity}
S=\frac{I_{PMT}/G(1800 V)}{P_{PD}}\frac{A_{PD}}{A_{PMT}}
\end{equation}
where $I_{PMT}$ is the current measured at the anode of the PMT in amps and $P_{PD}$ is the wavelength-corrected power measured by the calibrated photodiode in W.  Each was normalized to their respective active areas: $A_{PD}=1$~cm$^2$ and $A_{PMT}=20.26$~cm$^2$.  The resulting curve is found in \FIG \ref{fig:mono_results}. The maximum QE (@ 500~nm) was measured to be $(8.77\pm0.30)$\%.

\begin{figure}[t]
    \centering
\setupthecaption
    \includegraphics[width = \myfigurewidth \columnwidth]{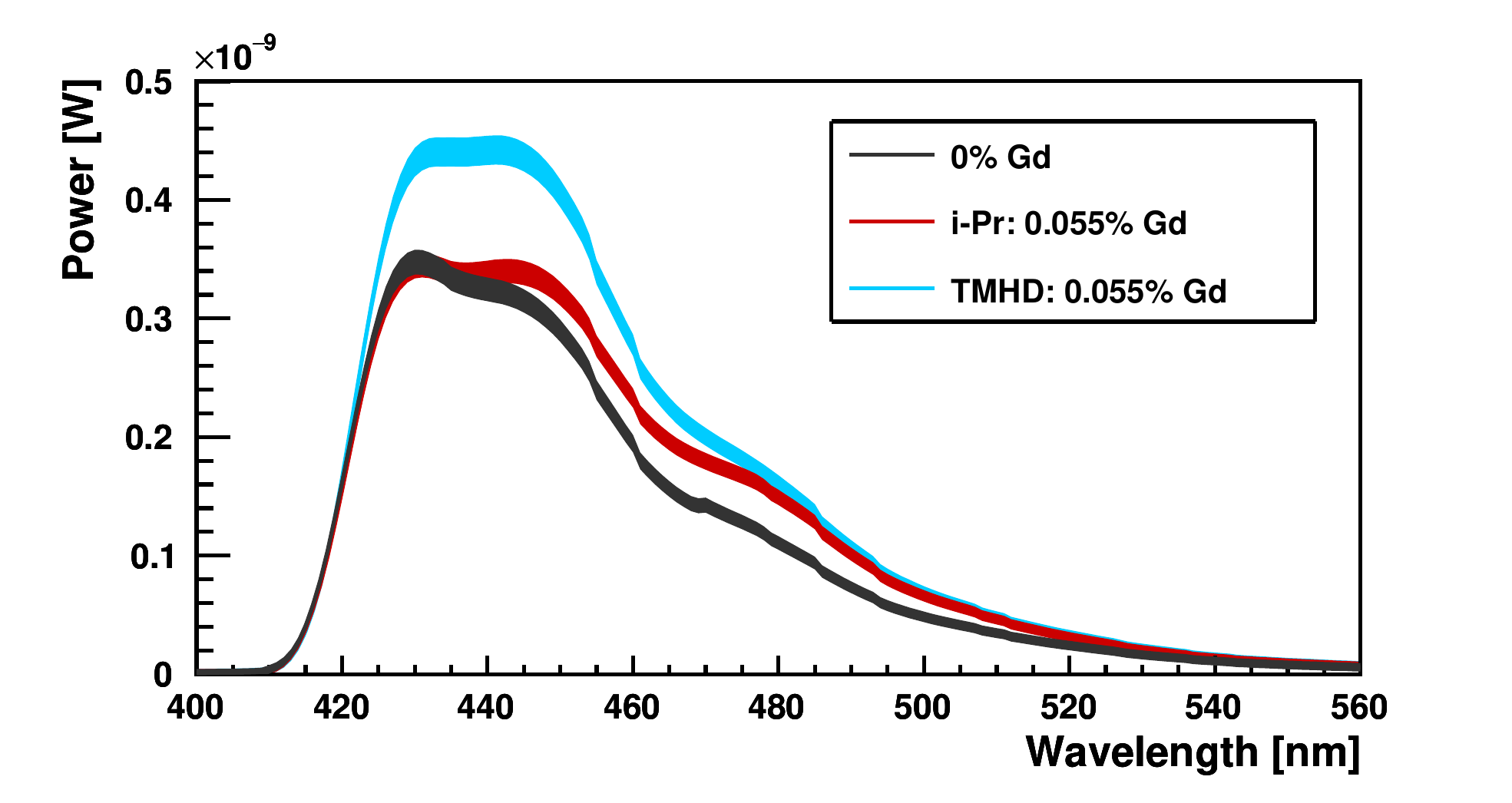}
    \caption{Emission spectra from unloaded and Gd-loaded samples measured in monochromator Configuration (b). The line width is representative of the systematic uncertainties derived from the standard deviation of repeated trials.}
	\label{fig:mono_results2}
\end{figure}

\subsection{Emission Spectra} \label{section:em_spectra}
The effect of different metal loadings on the relative light output and peak emission wavelength of the scintillator samples was measured using the monochromator setup shown in Configuration (b) of \FIG \ref{fig:MonoNew}.  The samples were all machined to the same length, originated from the same solution of fluors and styrene, and were placed in the same registered location relative to the monochrometer slit.  Each sample was wrapped in PTFE tape and backlit by a surface-mount LED of wavelength of 305~nm, chosen to correspond to the peak of the PPO absorption spectrum.    The LED at one end of the sample was offset from the direct line-of-sight presented by the monochromator slit at the other end.  The calibrated photodiode was used to measure the output power of the monochromator as a function of wavelength for each scintillator sample. 

The results of this test are shown in \FIG \ref{fig:mono_results2}. The peak emission wavelength for the Gd(i-Pr)$_3$ samples remained at $431.0\pm0.5$~nm, but an enhancement at ~445~nm can be seen at these low concentrations.  The TMHD(0.055\WT) showed a similar wavelength shift and also emitted more power over all wavelengths.  Enhanced light output after Gd-doping has been observed by the CEA group \CITE{dumazert2017gadolinium}. Ovechkina \textit{et al.\ }\CITE{Ovechkina} also reported an increase in the light output of Gd(i-Pr)$_3$ doped scintillator samples at concentrations higher than those reported in this paper.   

\subsection{Gd-doped Scintillator from an Industrial Partner}
\begin{figure}[t]
\centering
\setupthecaption
\begin{subfigure}[t]{0.495\columnwidth}  
	\centering
	\includegraphics[width = 0.8\columnwidth]{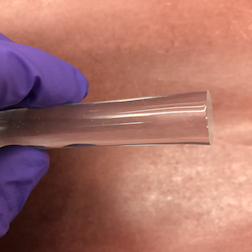}
	\caption{Small sample (2\WT Gd).}
	\label{fig:ind_sample1}
\end{subfigure}
\begin{subfigure}[t]{0.495\columnwidth}  
	\centering
	\includegraphics[width = 0.8\columnwidth]{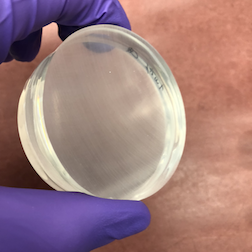}
	\caption{Large sample (1\WT Gd).}
	\label{fig:ind_sample2}
\end{subfigure}
\caption{Photographs of two samples produced by an industrial partner.  The cloudiness in (b) is only a surface effect and was easily polished away.  }
\label{fig:ind_samples}
\end{figure}

Any large-scale solid scintillator veto will need to be fabricated by an industrial partner with appropriate clean room capability.  In addition, high concentrations of uniform Gd-loading cannot be achieved without proprietary chemistry and our simulations have indicated that uniform loading is generally more efficient.  Thus, a collaboration was formed with an industrial partner who has successfully achieved loadings upwards of 2\WT Gd.  The CEA Saclay group~\footnote{CEA, LIST, Laboratoire Capteurs et Architectures Electroniques, CEA Saclay, 91191 Gif-sur-Yvette, France. E-mail: Guillaume.BERTRAND@cea.fr} provided clear samples with gadolinium loadings of up to 2\WT Gd.  We measured the emission spectra and light output as a function of Gd-loading for a set of four cylindrical samples that were all the same size (7.5 cm long and 1.85 cm diameter). Another four samples were larger in diameter, representing alternate casting styles.  Pictures of two of the samples are shown in \FIG \ref{fig:ind_samples}.

\begin{figure}[t]
	\setupthecaption
	\centering
	\includegraphics[width = \myfigurewidth \columnwidth]{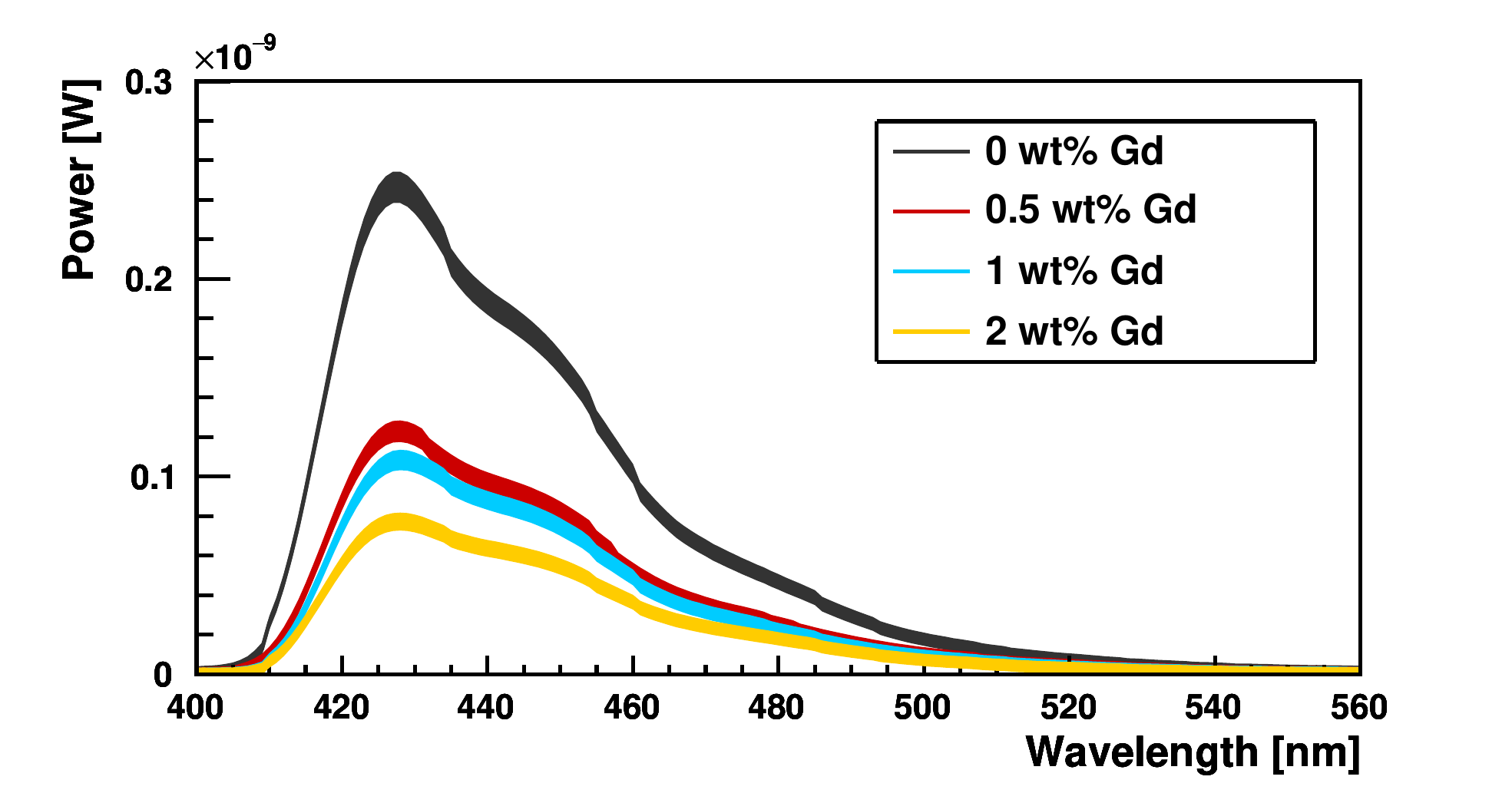} 
	\caption{The four small samples (see Figure~\ref{fig:ind_sample1} for an example) were measured in monochrometer configuration (b) to determine their emission spectra.  The line width of the emission spectra is representative of the systematic error determined by multiple trials.  }
	\label{fig:ind_spectra}
\end{figure}

Emission spectra were measured in the same manner as described in Section~\ref{section:em_spectra}. These results are displayed in \FIG \ref{fig:ind_spectra}. There is a noticeable increase in the green part of the spectrum with the addition of gadolinium, an effect also seen in our samples (See \FIG \ref{fig:mono_results2}). Our samples also showed an overall increase in light output, but only for loadings smaller than those available in the industrial samples.   By the time the Gd-loading reaches 0.5\WT, the light output at $427.0\pm0.5$~nm has dropped by almost 60\%, with a more gradual loss as the loading increases to 2.0\WT.  

Any veto element used in a low background dark matter experiment must be formed from highly radiopure materials. To understand if any of the chemical additives, fluors, or gadolinium compound contain contaminants that could compromise neutron veto functionality, one of the larger samples at 2\WT Gd-loading was assayed underground at SNOLAB in a coaxial high purity germanium (HPGe) detector.  This HPGe is used to do quality assurance on SuperCDMS tower and shield components, and is well-understood and thoroughly calibrated.   Results of this test are shown in Table~\ref{tab:industry_sample_radiopurity}. This test showed that samples are quite radiopure even at high gadolinium loadings, meaning that the polymerization process would not need to be altered significantly in scaling up to a full size neutron veto. 

\begin{table}[t]
\centering
\setupthecaption
\begin{tabular}{c|c}
Nuclide & Concentration (ppb) \\ \hline
 $^{238}$U from $^{226}$Ra &  $<0.13$ \\
 $^{238}$U from $^{234}$Th & $<4.79$ \\
 $^{235}$U & $1.85\pm6.38$ \\
 $^{232}$Th & $<0.17$ \\
 $^{40}$K & $<881.31$ \\
\end{tabular}
\caption{Sample screened in HPGe at the SNOLAB low background counting facility}
\label{tab:industry_sample_radiopurity}
\end{table}

\subsection{Attenuation Length Measurement} \label{section:atten_length}
The bulk attenuation length ($\lambda_{bulk}$) of a material can be simply defined as 
\begin{equation} \label{eqn:atten_length_eqn}
N(x) = N_0e^{-x/\lambda_{bulk}},
\end{equation}
where $x$ is the material thickness and $N$ is the number of photons.  This assumes that attenuation occurs due to absorption and that surface effects are negligible.  Commercially-produced plastic scintillators, such as BC-400, typically have bulk attenuation lengths of $\sim$25~m \CITE{SaintGobainSpecs}.  When measuring the attenuation length of small cylindrical plastic scintillator samples such as the ones fabricated, detailed optical effects (e.g. specular or diffuse reflectivity, index of refraction, sample geometry) become important.  We can define an effective attenuation length ($\lambda_{eff}$) as what is measured in the lab.  A full optical Monte Carlo simulation can be used to match the measured $\lambda_{eff}$  and thereby extract the optical properties such as reflectivity and bulk attenuation lengths.    
\begin{figure}[t]
\centering
\setupthecaption
\includegraphics[width = \myfigurewidth \columnwidth]{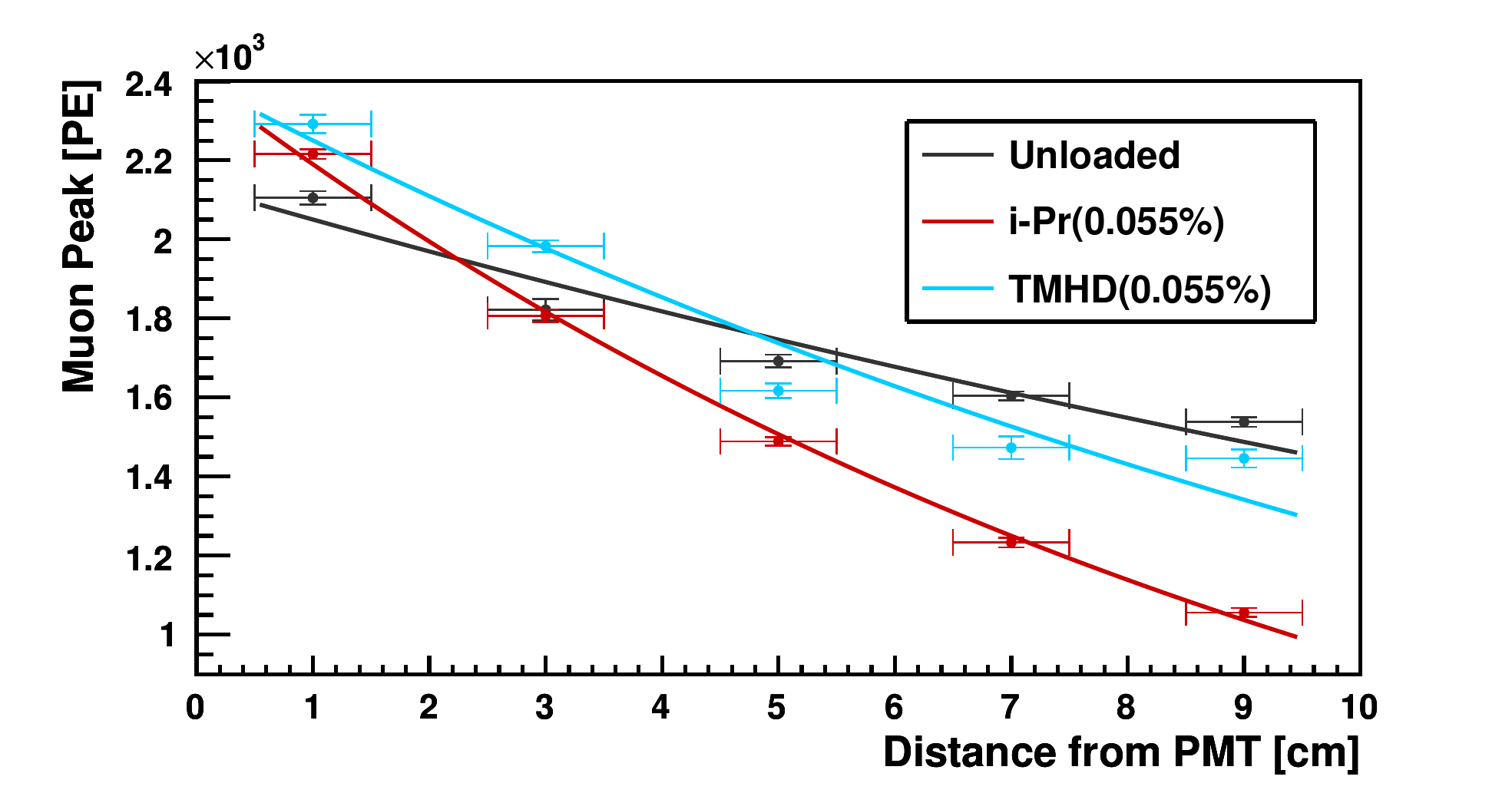}
\begin{tabular}{c|c|c}
& $\lambda_{eff}$ (cm) & $\chi_{red}^2$ \\ \hline 
Unloaded  & $24.9\pm2.1$ & $2.77$ \\ 
Gd(i-Pr)$_3$ (0.055\% Gd) & $10.70\pm0.85$ & $0.12$ \\  
Gd(TMHD)$_3$ (0.055\% Gd) & $15.5\pm1.3$ & $3.30$ \\  
\end{tabular}
\caption{The amount of light emitted from the end of the scintillator sample as a function of distance to the incident muon location, as determined by the center of the $2 \times 2$~cm$^2$ cross-section of the perpendicular trigger sticks.  Fits to a simple exponential are shown in the table. }
\label{fig:muontelescope}
\end{figure}

Since cosmogenic muons are close to minimum-ionizing, the mean energy that a muon will deposit in a scintillator sample depends only on the material density and muon path length, allowing for a reliable absolute calibration of the deposited energy.  A muon trigger was provided by the coincidence of two scintillator sticks of dimension 2~cm x 2~cm x 4~cm, oriented perpendicular to each other to create a 2~x~2~cm$^2$ area of  intersection.  The sample being tested was inserted between the two trigger sticks, wrapped in PTFE tape, and read out by the PMT digitized by a LeCroy Wavesurfer 434 oscilloscope.  A LabVIEW library was used to observe the waveforms and write data to text files. The area of each waveform in units of nV-s was converted to photoelectrons (PE). 

The sample scintillator remained in place, while the trigger sticks were moved the length of the sample in 2 cm increments.  Each run took 6-8 hours and collected $\sim$1000 events, resulting in a roughly Gaussian distribution with a sigma ranging from 10-15\% of the mean and small non-Gaussian tails from corner-clippers and high-angle muons. The mean and error on the mean for each Gaussian fit was plotted against distance to obtain the curves in \FIG \ref{fig:muontelescope}.  These data were fit to a simple exponential to obtain $\lambda_{eff}$.  At 1~cm, both loaded samples have light outputs greater than that of the unloaded sample, an effect of the enhanced light output from small concentrations of Gd-loading.  The light output of iPr(0.055\%) drops more rapidly than TMHD(0.055\%) as the length of the optical path increases.

\section{Strategies for Photodetection}

\subsection{Characterizing the WLS Fiber}
\begin{figure}[t]
\centering
\setupthecaption
\includegraphics[width = 0.5\columnwidth]{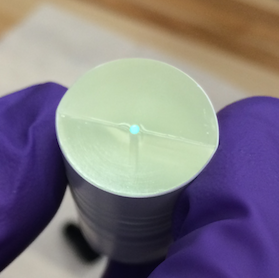}
\caption{Scintillator sample with WLS fiber embedded}
\label{fig:scint_with_fiber}
\end{figure}

A wavelength-shifting fiber (Kuraray CJ Y-11 (300)M) with a diameter of 1.00~mm was embedded in an unloaded sample by cutting the sample in half lengthwise and machining a groove down the center of both sides.  An attempt to avoid the machining step by adding the fiber during polymerization failed, due to the exothermicity of the polymerization reaction. 
After polishing, a thin layer of optical grease was applied to both halves of the machined sample. A fiber, cut to the length of the sample, was placed in the grove and the two samples halves were firmly pressed together to remove any air bubbles. The resulting samples had an elliptical cross section due to the machining. The sample was wrapped in several layers of PTFE tape.   The particular sample which was used in our trapping efficiency study was 4.5~cm long with major and minor axes of 2.11~cm  x 1.94~cm. It is shown in \FIG \ref{fig:scint_with_fiber}.  

\begin{figure}[t]
\centering
\setupthecaption
\includegraphics[width = \myfigurewidth \columnwidth]{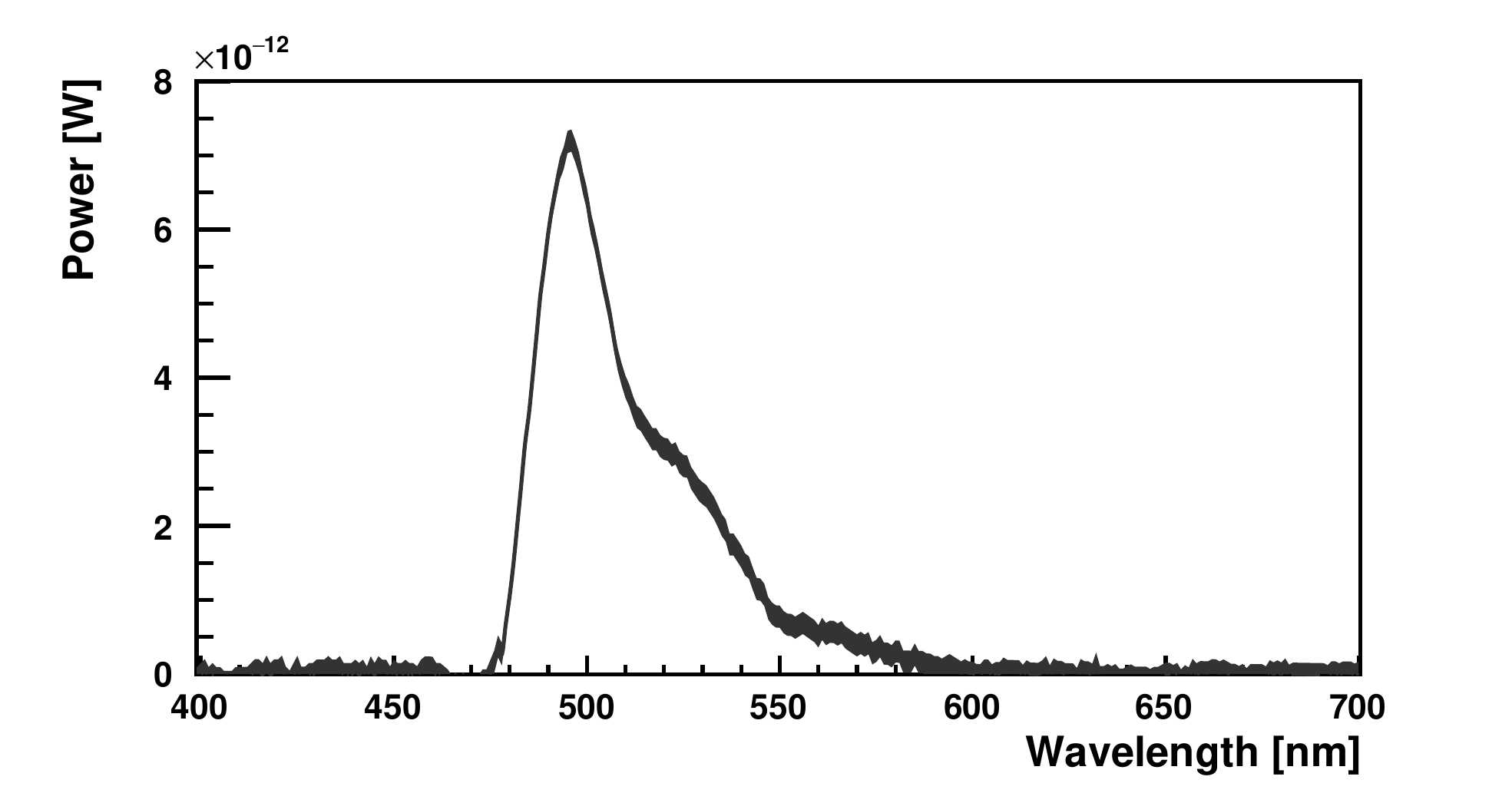} 
\caption{Measured WLS fiber emission spectrum. The maximum of the emission spectrum is located at ($496.0\pm0.5$)~nm. The line thickness is representative of the systematic error determined by multiple trials.}
\label{fig:WLS_emission}
\end{figure}

\begin{figure}[t]
\centering
\setupthecaption
\includegraphics[width = \myfigurewidth \columnwidth]{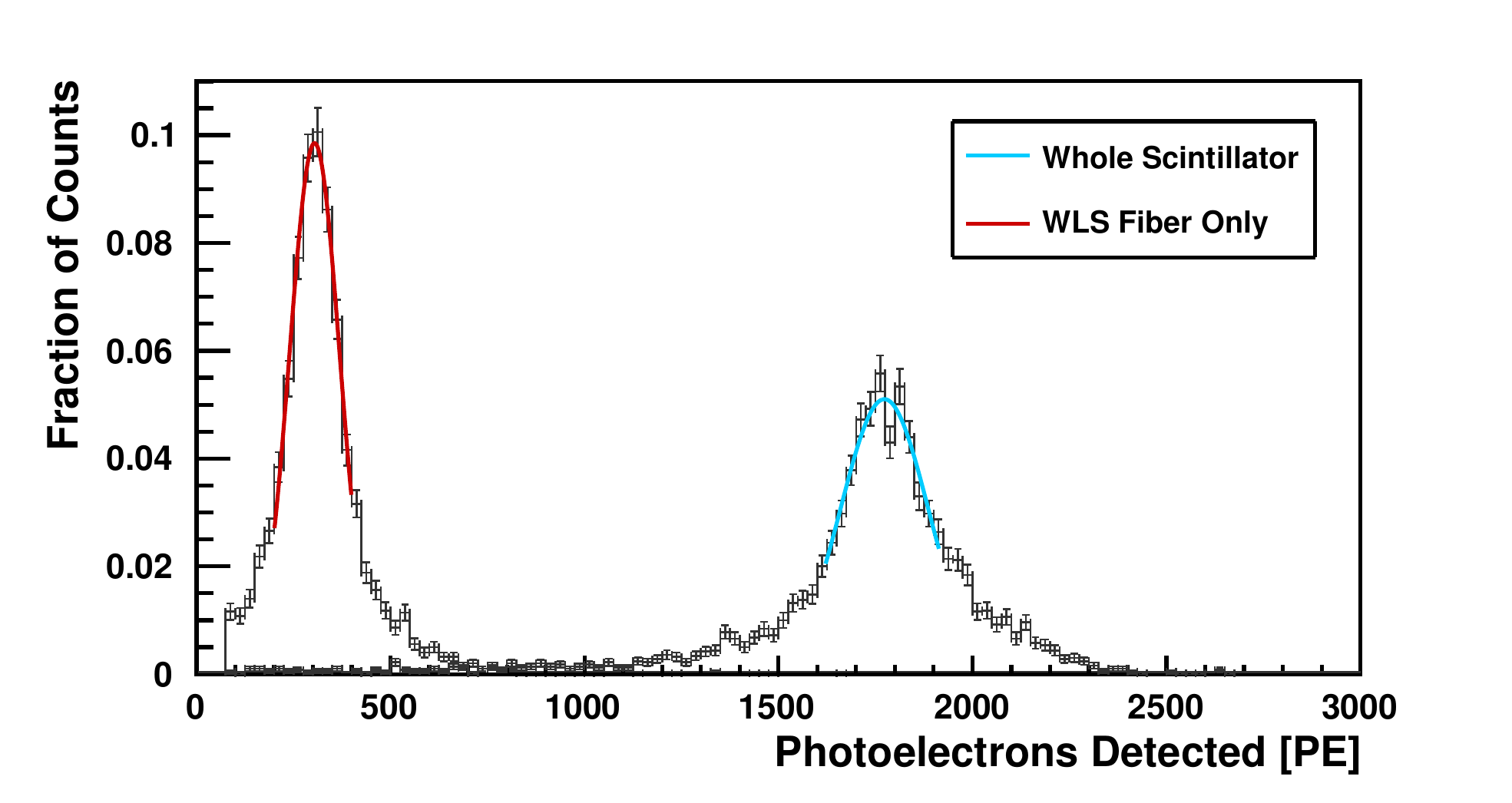} 
	\caption{The number of photoelectrons detected by the PMT from scintillator and fiber together, compared to the number detected from the fiber through a foil mask.}
	\label{fig:WLS_trapping}
\end{figure}

To measure the emission spectrum of the fiber, a white light source was used to illuminate the WLS fiber, shown in Configuration (c) of \FIG \ref{fig:MonoNew}. A black cardboard shield with a 1~mm diameter fiber feedthrough was used to block light from the source.  Prior to fiber insertion, monochromator runs determined that the power measured with the source on was within $2\sigma$ across all wavelengths of the power measured in total darkness.  \FIG \ref{fig:WLS_emission} shows the measured emission spectrum of the fiber. This matches the manufacturer's spectrum given by Kuraray~\CITE{kurarayY11}.

\subsection{Photon trapping by the WLS fiber}\label{section:fibertrapdata}

The trapping efficiency of the fiber can be defined as the fraction of 
optical photons which enter the fiber's volume, are wavelength-shifted, and reach the face of the photodetector, normalized to those generated in the main scintillator per incident muon.   This can be determined by a Monte Carlo simulation (See Section~\ref{section:Monte_Carlo}).  We can define an experimentally accessible quantity, $\eta$, called the "detection efficiency" as: 
$\eta = \frac{PE_{WLS}}{PE_{tot}}$, where $PE_{tot}$ is the number of photoelectrons detected by the PMT from the entire face of the sample and $PE_{WLS}$ is the number detected by the same PMT viewing only the WLS fiber.   This measured quantity of $\eta$ can be used to tune the simulation and thereby extract the more theoretical trapping efficiency, 

The detection efficiency was measured using the muon telescope positioned 2~cm away from the PMT face.  The light from the entire sample was observed by the PMT directly coupled to its end.   Then a foil mask was inserted, which restricted the light to that from the fiber.   The two muon-induced peaks are shown superposed in \FIG \ref{fig:WLS_trapping}.  Gaussian fits gave values of ($1770\pm110$)~PE for the scintillator and fiber together, and ($304\pm64$)~PE for the fiber surrounded by a foil mask, yielding a detection efficiency of 17\%.  Since the hole in the aluminum mask was larger than the fiber, the number of PE measured from the WLS fiber included a 0.5 mm wide annulus of blue light directly from the scintillator.

The advantage conferred by concentrating the light in a fiber is better appreciated by taking the ratio of photons per unit area.  This "enhancement factor" is $\eta_c$:
\begin{equation} \label{eqn:WLS_concentration}
\eta_c=\eta\times\frac{\int QE_{scint}}{\int QE_{WLS}}\times\frac{A_{scint}}{A_{hole}}
\end{equation}
and is obtained by correcting $\eta$ by the area of scintillator ($A_{scint}= 321.5 mm^2$) and hole in the mask  ($A_{hole} = 3.14 mm^2$), as well as the ratio of the convolution of the photodetector QE with the measured emission spectra of the blue scintillator versus the green fiber ( $\int QE_{scint}/\int QE_{WLS} = 0.95$).   For the sample with the embedded WLS fiber, the photon flux per unit area is enhanced by a factor of $\eta_c$ = $16.7\pm3.7$.

\subsection{Silicon Photomultiplier Readout}
The photodetector of choice for a dark matter experiment using a solid scintillator neutron veto is a silicon photomultiplier (SiPM).  This is due to its much lower radioactivity.   Since a SiPM is an array of tiny avalanche photodiodes (the pixels), the photon detection efficiency (PDE) is a product of the effective area coverage (accounting for the dead space between pixels) and the probability of an avalanche above threshold for the energy deposited.   A typical SiPM with a pixel size of 1~mm$^2$ can have a maximum PDE of around 30-70\%~\CITE{SiPM_hamamatsu}. 

The PDE of a SiPM can be measured by dividing the rate of photoelectrons detected per second ($PE_{detected}$) by the rate of photons incident on the SiPM's active area per second ($n_{incident}$):
\begin{equation}
\label{eqn:SiPM_PDE_eqn1}
\text{PDE}(\lambda) = \frac{PE_{detected}}{n_{incident}}.
\end{equation}
Since the power per unit area of a light source ($I$) at a specific wavelength was measured by the calibrated photodiode, the value of $n_{incident}$ can be calculated as
\begin{equation}
\label{eqn:SiPM_PDE_eqn2}
n_{incident} = \frac{IA_{\mathit{SiPM}}}{hc/\lambda},
\end{equation}
where $A_{\mathit{SiPM}}$ is the active area of the SiPM~\CITE{SiPM_hamamatsu}.

\begin{figure}[t]
\centering
\setupthecaption
	\includegraphics[width = \myfigurewidth \columnwidth]{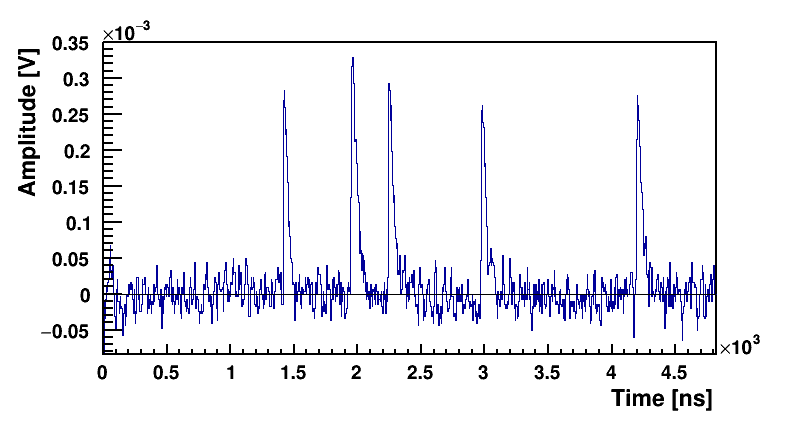}	
	\caption{Example of a long SiPM waveform used to find the dark count. }
	\label{fig:sipm_waveform_long2}
\end{figure}

\begin{figure}[t]
\centering
\setupthecaption
	\includegraphics[width = \myfigurewidth \columnwidth]{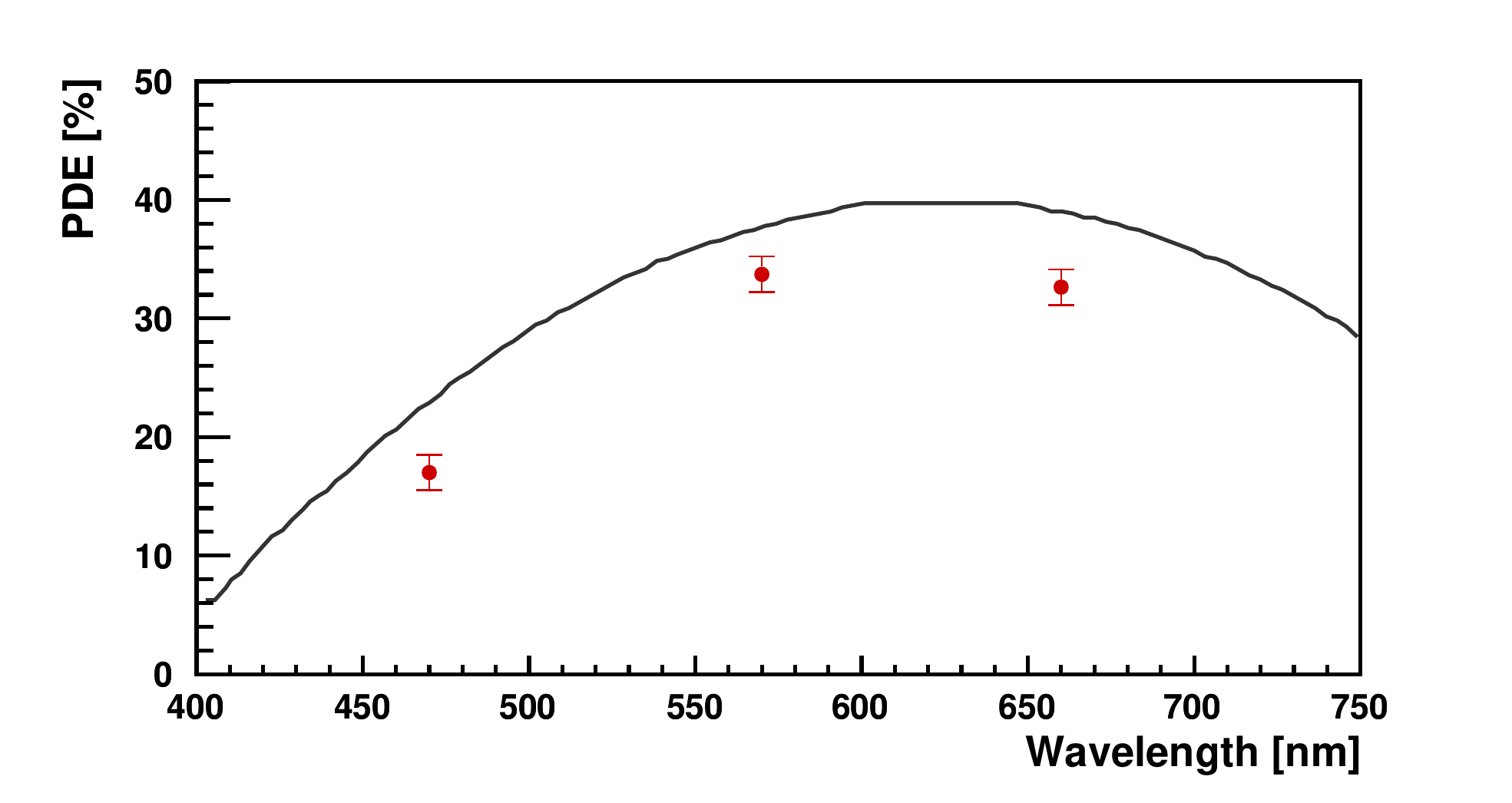}	
	\caption{Measured photon detection efficiency (PDE) of the CPTA 151 SiPM (red squares) compared to the manufacturer's standardized curve. \CITE{cptaSpecs}.}
	\label{fig:sipm_pde_results}
\end{figure}

The value of $PE_{detected}$ was measured by actually counting the number of pulses over a fixed time period while the SiPM was illuminated by the light source.  The dark count rate was measured in the same way while the light was off, and then subtracted.  In practice, this was done by integrating over a long waveform, an example of which is shown in \FIG \ref{fig:sipm_waveform_long2}, to determine the total number of photoelectrons detected per unit time.  The experimental setup is shown in \FIG \ref{fig:MonoNew} Configuration (a) with a  SiPM biased to 20.7~V.  The average SiPM gain was $2.25 \times 10^6$.  To avoid saturation, pileup, and SiPM bias drift, a low-intensity light source supplied by a DC voltage was used for this measurement.  The measured PDE values for one of our SiPMs is shown in  \FIG \ref{fig:sipm_pde_results} and closely resembles the manufacturer's spectrum (overlaid). 

\begin{figure}[t]
\centering
\setupthecaption
\includegraphics[width = \myfigurewidth \columnwidth]{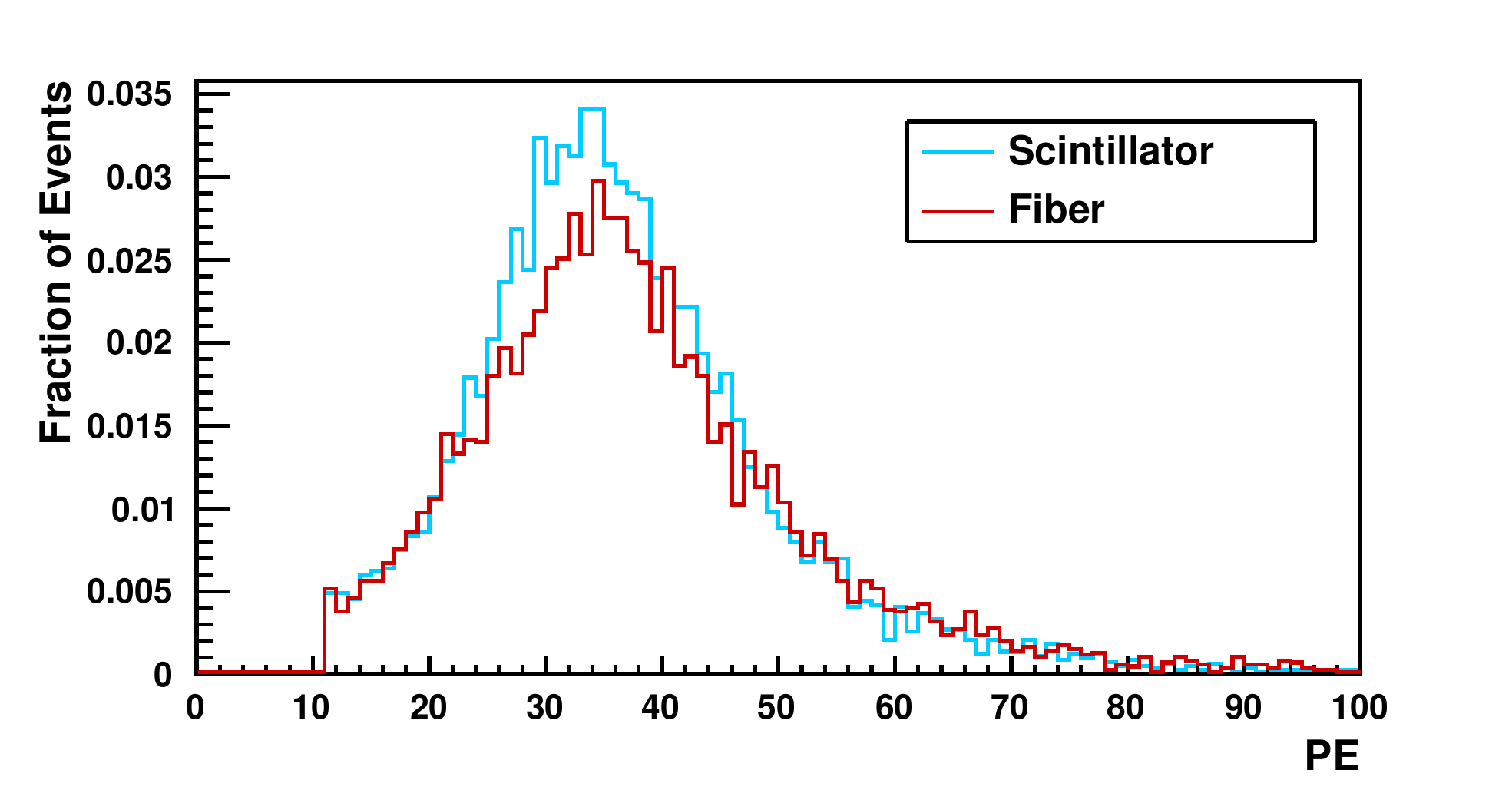}
\caption{The number of PE per incident muon as read out by one SiPM on the end of the WLS fiber compared to the sum of three SiPM's reading out the blue scintillator.  The average number of PE were determined by fitting Gaussians to the distributions.}
\label{fig:sipm_eta_results}
\end{figure}
A second measurement of the enhancement factor $\eta_c$ was made with the muon telescope trigger, using SiPMs in place of the PMT.   Two Teflon caps were machined to covered the readout end of the sample.  Data from the WLS fiber was read out by a SiPM in the center of the cap and gave an average number of  $n_{WLS}=34.4\pm9.9$~PE per incident muon.  Data from the blue scintillator was read out by a set of three SiPMs arranged in a triangle around the center.  Each SiPM had a different gain and overall PDE, which was applied individually in order to convert them to PE.   The corrected PE sum of the SiPMs on the scintillator, divided by 3 to normalize to the same area as the SiPM reading out the WLS fiber, gives $n_{scint}=11.4\pm3.2$~PE.  Thus, $\eta_c$, here defined as
\begin{equation} \label{eqn:WLS_concentration2}
\eta_c = \frac{n_{WLS}}{n_{scint}}\times\frac{\int PDE_{scint}}{\int PDE_{WLS}},
\end{equation}
gives an enhancement factor of only $2.9\pm1.5$, compared the factor of almost 17 observed by the PMT measurement.  The reasons for this are described in detail in Section~\ref{section:Monte_Carlo}.  Briefly, there are two major effects:  (1) the low-reflectivity teflon cap which masked the entire scintillator face for both $n_{scint}$ and $n_{WLS}$ measurements is responsible for a factor of 3 reduction, and (2) the hole in the aluminum mask used for the PMT measurement let in additional blue light, yielding another factor of 1.8.  Applying both correction factors to the SiPM measurement gives an enhancement factor of $\eta_c$ of 16, consistent with the PMT measurement.  It should also be noted that the simulation shows that many of the trapped optical photons are propagating near the outer edge of the fiber, making the light output from the fiber very sensitive to any misalignment of a SiPM sensor of the same size.

\section{WLS Fiber Trapping Efficiency} \label{section:Monte_Carlo}

\subsection{\geant-based Simulations of Plastic Scintillator}

In order to determine the fiber trapping efficiency of our Gd-loaded plastic scintillator samples, the experimental setup was modeled using \geant v10.02. \geant is a C++ based toolkit~\CITE{AGOSTINELLI2003250}, which simulates the passage of particles through matter by tracking individual particle interactions and recording their energy deposition at each step through a material. To handle hadronic and electromagnetic interactions, the modules G4ShieldingPhysicsList and G4EMPhysicsList\_option4 were employed. The G4OpticalPhysicsList and the UNIFIED Optical Boundary model were used to handle optical interactions and boundary crossings, respectively.  The absorption spectra for the scintillator cocktail and for the green WLS fiber were taken from literature (see \FIG \ref{fig:fluorSpectra}). However,  for the emission spectra of the scintillator and fiber, and for the PMT quantum efficiency we used our own monochrometer measurements     (\FIGS \ref{fig:mono_results} and \ref{fig:WLS_emission}). Other properties, notably the reflectivity and attenuation lengths of our samples, were inferred from tuning the simulation to the experimental data on $\lambda_{eff}$. Plastic scintillator was defined as a composition of predefined elements:  52.4\% hydrogen and 47.6\% carbon, yielding an H:C ratio of 1.100 and a density $\rho_{scint}=1.05 g/cm^3$, similar to the specifications stated by the manufacturer \CITE{SaintGobainSpecs}. In the scintillator material definition, Birk's constant~\CITE{Birk}
         was set to 0.149 mm/MeV .

The muon telescope data was triggered by the coincidence of crossed scintillator sticks placed above and below the sample. The trigger was modeled in \geant by generating muons of 4 GeV (corresponding to the average energy of cosmogenic muons at sea level) uniformly across the entire top scintillator stick with the appropriate zenith angle distribution downward.   Only those muons which also passed through the second crossed scintillator stick below the sample were tracked through the rest of the simulation.  As a check, a simulation with the full kinetic energy spectrum expected for muons at sea level was run for one location and no difference in the final light output was observed. Two sample geometries were modeled: the unloaded sample (11 cm long $\times$  2.5 cm diameter) and the sample with a WLS fiber down the center (4.5 cm long $\times$ $\sim$2 cm diameter elliptical).  

Optical photons with the measured emission spectrum were generated by the muon interactions in the scintillator.  The index of refraction of the scintillator was set to 1.58, corresponding to the data sheet value given by the manufacturer \CITE{SaintGobainSpecs}. Their reflections were handled by the UNIFIED optical reflection model~\CITE{Levin}. To simulate the reflective wrapping on walls and ends of samples, PTFE was designated as a dielectric material producing diffuse reflections with a simple absolute value $R_{PTFE}$ corresponding to the probability of reflection at the  boundary.  The value of $R_{PTFE}$ was tuned by matching to data. 

For WLS fiber light collection studies,  the following indices of refraction were used for the double-clad Kuraray Y11 fiber \CITE{kurarayY11}: $n_{core}=1.59$, $n_{innerClad}=1.49$, $n_{outerClad}=1.42$. Optical photons that crossed into the WLS fiber volume were absorbed and re-emitted into 4$\pi$ with the appropriate absorption and emission spectra.  Any optical photon that crossed the PMT boundary was converted to a photoelectron based on the measured quantum efficiency of the PMT.   All photon tracks that were not converted were killed.

\begin{figure}
	\centering
\setupthecaption
	\includegraphics[width=0.9\columnwidth]{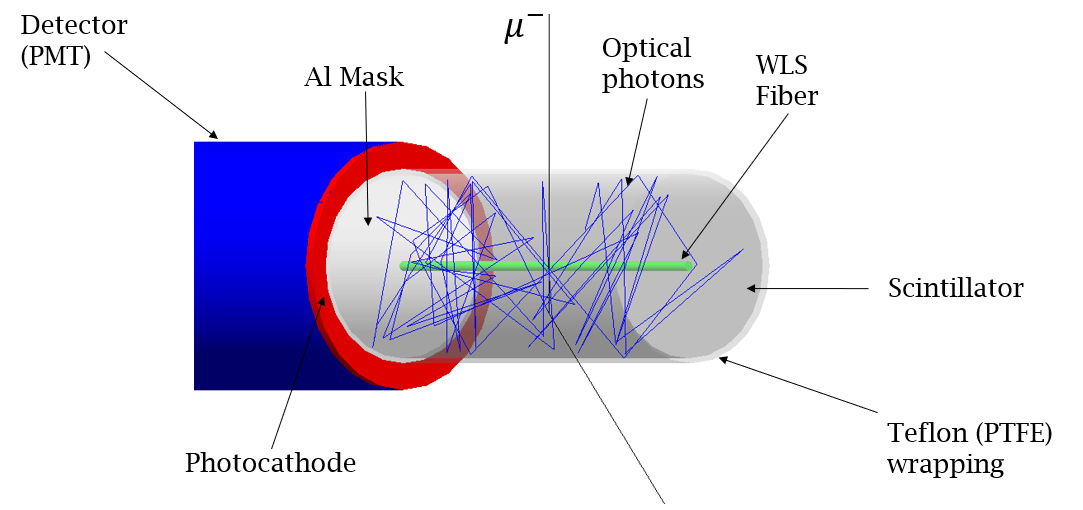}
	\caption{\geant geometry of the WLS fiber sample illuminated by a muon triggered by the crossed scintillator sticks (not shown).  The Al mask and WLS fiber were only present for simulations where we wanted to observe only the light collection properties of the WLS fiber. For attenuation length measurements, the sample was longer and the mask and fiber were not present.}
	\label{fig:simulation_geometry}
\end{figure}

\begin{figure}[h]
\centering
\setupthecaption
	\includegraphics[width=.49\columnwidth]{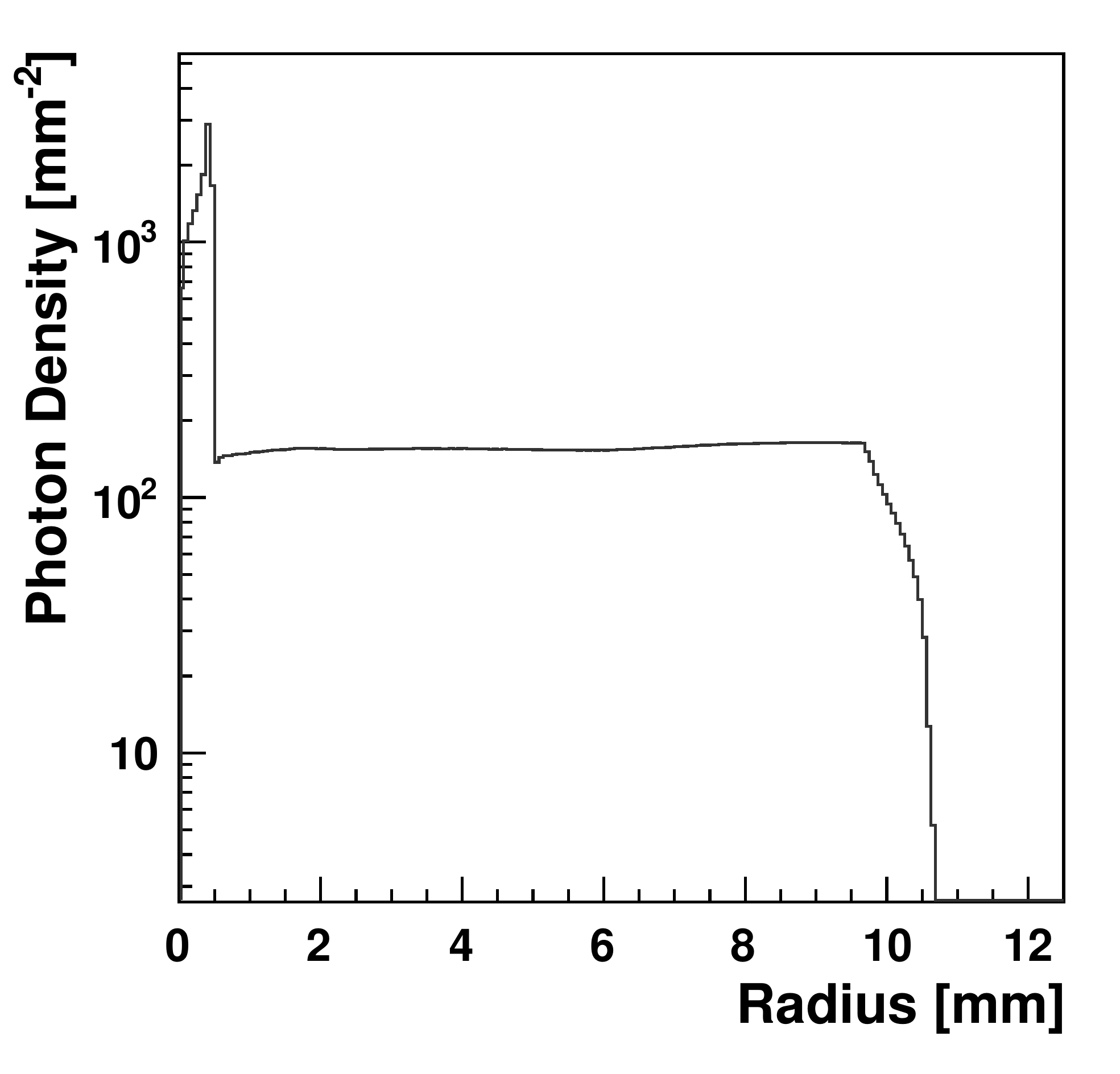}
	\includegraphics[width=.49\columnwidth]{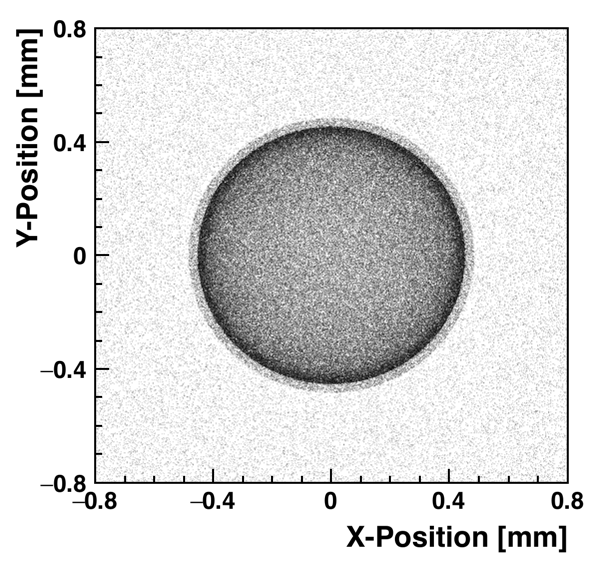}
\caption{\geant results for the scintillator sample with embedded WLS fiber.   Left: the density of the impinging photons on the PMT face as a function of radius, normalized to the area of the differential ring corresponding to that radius.  Right: An x-y plot of the PMT impact position of the photons for the same sample.   }
\label{fig:pmtXYLight}
\end{figure}

A muon-induced event generated by \geant is shown in \FIG \ref{fig:simulation_geometry}, also illustrating our simulation geometry.  The light trapping properties of the fiber are shown in \FIG \ref{fig:pmtXYLight}.  Note that while illumination from the scintillator is fairly uniform, the trapped light is concentrated near the periphery of the fiber.  The cladding is also visible as a lighter outer ring in \FIG \ref{fig:pmtXYLight}b, contributing somewhat to the trapping. 

Even more light can be collected in the central fiber if the PMT photocathode face surrounding the fiber is covered with a reflective aluminum foil mask. The foil mask was represented in the simulation by a layer of aluminum of thickness 20 {$\mu$}m undergoing specular reflection with an absolute reflectivity $R_{Al}$.  The reflectivity of the foil was varied in order to understand the degree to which it can affect fiber trapping efficiency.  The results of this study are shown in \FIG \ref{fig:AlStudy}  indicating that end cap reflectivity is crucial to designing the most efficient scintillating-fiber readout. 

The effect of a small air gap between the scintillator and the PTFE wrapping was also studied. The size of the air gap was not measurable, but it was not zero, as confirmed by the presence of total internal reflection in wrapped samples.  For simulations where the end of the sample mated directly to a PMT, the presence of an air gap and its thickness (up to 200 $\mu$m) did not affect fiber trapping results.  However, when a highly reflective Al mask was included, the signal from the fiber increased by 30\% as the gap was increased from 0 to 100 $\mu$m, falling off again for larger gaps.  The simulation results described below all included an air gap of thickness 100 $\mu$m.  

\subsection{Constraining reflectivity and bulk attenuation length}

\begin{figure}[t] 
	\centering
\setupthecaption
	\includegraphics[width = \myfigurewidth \columnwidth]{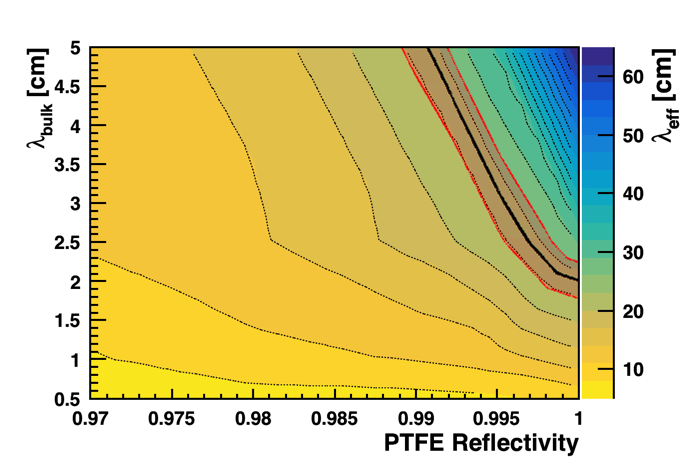}
	\caption{Contour Plot visualizing the data from Table \ref{tab:effAttLengTab}. The experimental value for $\lambda_{eff}$ and its one sigma envelope is represented by the black contour and red shaded region.}
	\label{fig:GraphTableData_Unmasked}
\end{figure}

The surface reflectivity $R_{PTFE}$ (depends on wrapping, air gap, and surface qualities) and the bulk attenuation (depends on plastic and fluor) of the unloaded scintillator sample can be determined by comparing simulation to $\lambda_{eff}$ measurements.   Simulations were performed for muons intersecting the scintillator sample centered at a position $d=1,3,5,7,9$ cm away from the PMT face. Literature defined a reasonable range about which to vary the reflectivity \CITE{ptfeReflArticle} and the bulk attenuation length~\CITE{SaintGobainSpecs}. The dependence of $\lambda_{eff}$ on reflectivity and bulk attenuation is shown in Table \ref{tab:effAttLengTab} and plotted in \FIG \ref{fig:GraphTableData_Unmasked}. For reflectivities above 95\%,  $\lambda_{eff}$ is highly dependent on the reflectivity of the outer wrapping as long as the bulk attenuation is larger than a meter.  It is clear from the 2-D contour that there are multiple solutions which match the data.  Since the bulk attenuation length of undoped scintillator is quoted by Saint Gobain~\CITE{SaintGobainSpecs} to be  $\lambda_{bulk}=$2.5 m, we can fix that value and choose the outer wrapping reflectivity to be $R_{PTFE}=0.995$ in order to match our measured effective attenuation length, for purposes of finding the trapping efficiency. 

\begin{table}[t]
\centering
\setupthecaption
\resizebox{\columnwidth}{!}{%
\begin{tabular}{@{} c|p{3.3}|p{4.3}|p{4.3}|p{4.3} @{}}
\centering
PTFE & \multicolumn{4}{c}{$\lambda_{eff}$ (cm)} \\ \cline{2-5} %\multirow{2}{*}{PTFE Reflectivity}
Reflectivity & \multicolumn{1}{c|}{$\lambda_{bulk} =$ 0.5 m} & \multicolumn{1}{c|}{$\lambda_{bulk} =$ 1.0 m} & \multicolumn{1}{c|}{$\lambda_{bulk} =$ 2.5 m} & \multicolumn{1}{c}{$\lambda_{bulk} =$ 5.0 m} \\ \hline
1.00 & 8.9,0.1 & 15.8,0.4 & 29.3,1.1 & 66.5,6.4\\
0.995 & \multicolumn{1}{c|}{---} & \multicolumn{1}{c|}{---} & 22.1,0.4 & 32.5,0.6\\
0.99 & 7.1,0.1 & 11.2,0.2 & 18.0,0.4 & 23.6,0.6\\
0.98 & 6.8,0.1 & 10.1,0.2 & 13.5,0.2 & 14.7,0.3\\
0.97 & 6.7,0.1 & 7.7,0.1 & 11.4,0.3 & 12.9,0.2\\
\end{tabular}%
}
\caption{Monte Carlo results for $\lambda_{eff}$ from an exponential fit to the number of PE as a function of the distance to the incident muon position.}
\label{tab:effAttLengTab}
\end{table}

\subsection{Modeling the WLS Fiber}

\begin{figure}[t]
    \centering
\setupthecaption
    \includegraphics[width = \myfigurewidth \columnwidth]{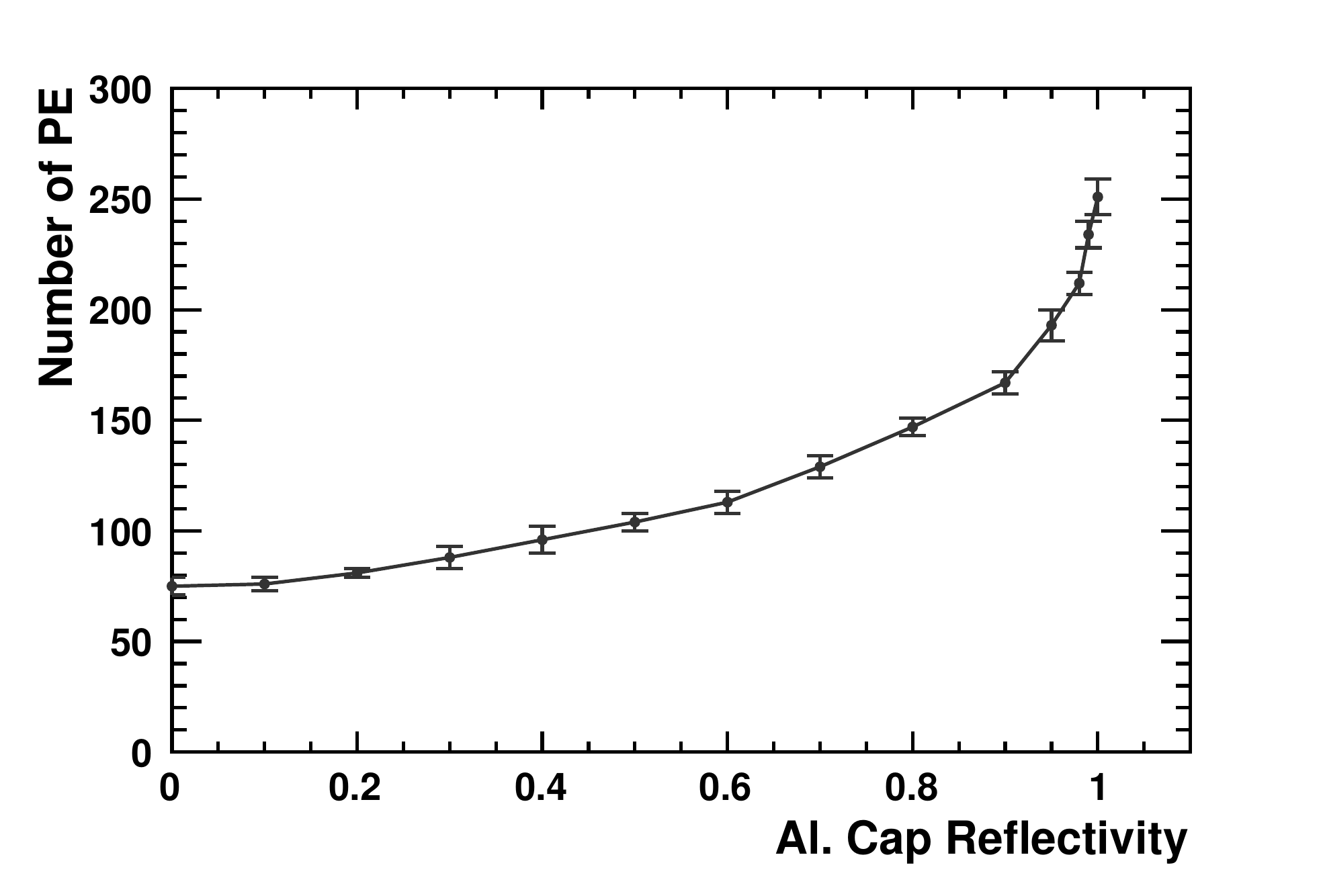}
    \caption{The number of WLS fiber PE detected by a PMT as a function of the reflectivity of aluminum mask surrounding the fiber.   In this case, the hole in the mask is well-matched to the diameter of the fiber. }
    \label{fig:AlStudy}
\end{figure}

In order to determine the trapping efficiency, defined as the fraction of muon-induced optical photons that make it to the PMT via the WLS fiber, a simulation with the embedded WLS fiber geometry was performed.    In order to compare with the data shown in Figure~\ref{fig:WLS_trapping}, muon tracks were generated on the crossed trigger sticks located 2 cm down the axis of the 4.5 cm long sample. Simulations were performed for the case where the PMT was illuminated by the full surface area of the scintillator, as well as masked by a reflective end cap with a hole of varying diameter between 1.05 mm to 2.40 mm.  Aluminum reflectivity was varied from 0 to 1. The average number of PE detected by the PMT were determined by fitting to Gaussians with uncertainties corresponding to the sigma of the fit.    For runs where the scintillator was not masked, the average number of photoelectrons detected were $79\pm4$ PE from the fiber and $2149\pm90$ PE from the entire face.  For runs which included a reflective end cap, the WLS fiber signal increased as a function of the aluminum reflectivity, as shown in \FIG \ref{fig:AlStudy} for a close-fitting 1.05 diameter hole.  

In order to match the measured detection efficiency of $\eta = \frac{PE_{WLS}}{PE_{tot}}$ = 17\% from Section~\ref{section:fibertrapdata}, the numerator must correspond to a highly-reflective aluminum end cap (which triples the WLS fiber signal), while the denominator corresponds to the unmasked case.  Expanding the size of the hole also increases the number of PE.  The hole in our aluminum mask was not perfectly circular, with an average diameter of 2 mm.  The simulation matches the data best for $R_{Al}$ = 0.98,  D$_{hole}$= 2.0 mm, but is also marginally consistent with $R_{Al}$ = 0.99,  D$_{hole}$= 1.8 mm.  Assigning a reflectivity of 0.98 to the aluminum endcap,  the detection efficiency expected from a fiber with a perfectly matched hole would be $\eta$=10.4\%, and the photons per unit area emitted from the fiber would be larger by a factor ($\eta_c$) of 37 than the photons per unit area from the entire face. 

Having determined a set of optical parameters which match the data, the simulation can be used to calculate the theoretical trapping fraction for our sample configuration.   Setting $R_{Al}$=0.98 for the mask reflectivity and using the bulk attenuation and PTFE reflectivity (plus gap) established above, the trapping efficiency is $9.3\%$.  

It is useful to compare these results to a simple analytical calculation of the trapping probability for photons generated into a 4$\pi$ solid angle in the center of a 1~mm diameter fiber which is double clad with the same set of refractive indices as the Kuraray Y11.  Following Papandreou \CITE{fiberTrappingCalc}, we get a value of  10\%.  This simple treatment corresponds to the case where 100\% of those photons which enter the fiber get converted into WLS photons and then travel through the fiber without absorption, thus over-estimating the trapping fraction.  This is partially compensated by the fact that the calculation ignores sidewall reflectivity, thus under-estimating the number of chances that a photon can enter the fiber.\\

\section{SuperCDMS SNOLAB Neutron Veto Design}

\subsection{The SuperCDMS SNOLAB Simulation}

\begin{figure*}[h]
\centering
\begin{subfigure}{0.495\textwidth}  
	\centering
	\includegraphics[width = 0.65\columnwidth]{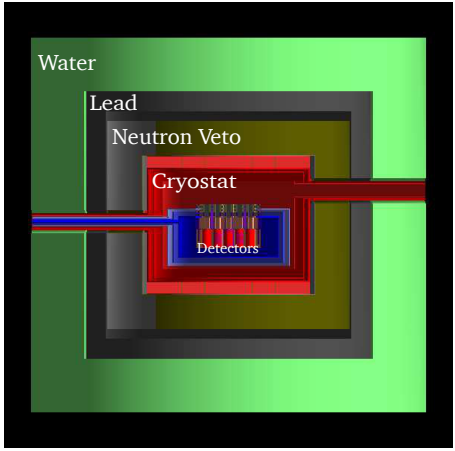}
	\end{subfigure}
\begin{subfigure}{0.495\textwidth}  
	\centering
	\includegraphics[width = 0.65\columnwidth]{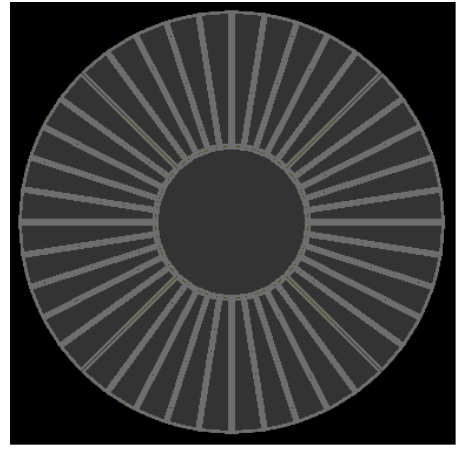}
	\end{subfigure}
\caption{Left: Cross section of the SuperCDMS SNOLAB shield and cryostat geometry in the \geant Monte Carlo. The neutron veto is a cylinder surrounding the cryostat. Right: Top view of the proposed active neutron veto which would replace the current passive polyethylene shield.  }
\label{fig:veto_pic}
\end{figure*}

For the first 4-tower payload of the SuperCDMS SNOLAB dark matter experiment~\footnote{https://supercdms.slac.stanford.edu}, the neutron background will be negligible; radiogenic neutrons from the cavern environment will be blocked by the shielding and higher-energy muon-induced neutrons are very rare due to the depth (2~km) of the laboratory.    However, for a larger payload and longer exposure a small number of single-scatter, neutron-induced nuclear recoil events are expected to be observed in the detectors.  Some of these neutrons are due to trace contaminants in the shield itself, as well as from the rare muon that makes it to the SNOLAB depth. 
A diagram of the shield is found on the left panel of \FIG \ref{fig:veto_pic}.  The following \geant simulation study, using the detailed SuperCDMS SNOLAB shield and detector geometry, was performed for a 25-tower payload,  in order to understand if an efficient neutron veto could be created by replacing the passive polyethylene layer of the shield with scintillator doped with gadolinium.  

The primary neutrons were thrown from the inner layer of the cryostat with an energy spectrum corresponding to the equilibrium thorium chain since the thorium spectrum is similar to other potential contaminants and the inner cryostat contributes the largest fraction of neutrons that make it to the detectors.  The active neutron veto simulated was composed of wedge-shaped sections of plastic scintillator with embedded wavelength-shifting fibers read out by SiPMs. These wedges formed a 40.6~cm thick cylinder with an inner radius at 98.4~cm, as shown in the right panel of \FIG \ref{fig:veto_pic}. 

\subsection{Optimizing Veto Efficiency} \label{section:veto_efficiency}

The simulation  compared uniformly-doped plastic to configurations where undoped plastic scintillator was interleaved with non-transparent layers of highly-concentrated gadolinium (Gd).   This can be achieved with over-loading each scintillator layer and orienting the undissolved Gd along the bottom surface of each wedge or by coating undoped scintillators with a Gd resin.  The latter option was modeled in \geant, using the following combination of compounds developed by the ZEPLIN-II experiment: 2.76 kg of gadolinium nitrate hexahydrate (Gd(NO$_3$)$_3 6$H$_2$O) combined with 1.92 kg of styrene monomer \CITE{hassan}. 

Each of the wedge-shaped segments was coated with Gd resin, represented by the light grey layers in the right panel of \FIG \ref{fig:veto_pic}, which is an example of a veto design with 40 segments. In order to represent an exact replacement of the passive polyethylene shield layer, the active neutron veto layer radial thickness remained fixed at 40.64~cm.  The thickness of the resin was set such that the percent weight of Gd included in the total mass of the resin remained equivalent to the total amount of Gd in the uniformly-doped scintillator option. Thus, the more finely segmented the cylinder, the thinner the resin layer around each wedge.  Thus segmentation cost would be scaled by the mechanical design rather than the quantity of gadolinium. 

The efficiency of the active neutron veto was defined with respect to events which deposit energy above threshold in only one detector (a single-scatter nuclear recoil). The efficiency is thus the fraction of those single events which then scatter back into the neutron veto and deposit energy above threshold in the scintillator within an integration time.  The final energy threshold will be determined by the light collection efficiency and photodetector threshold.  The process of nuclear capture and the subsequent release of capture gammas takes hundreds of microseconds, so collecting all the light may require up to a millisecond.  However, during that time, dark current and electronic noise will be collected along with any signal, making integration time a variable which must be tuned depending on desired dead time and the expected background rate.   Only the energy deposition as a function of time is presented here.  The simulation did not track optical photons or photo-detection efficiency.  

\begin{figure*}[t] %{0.9\textwidth}
\centering
\begin{subfigure}{0.495\textwidth}  
	\centering
	\includegraphics[width = 0.95\columnwidth]{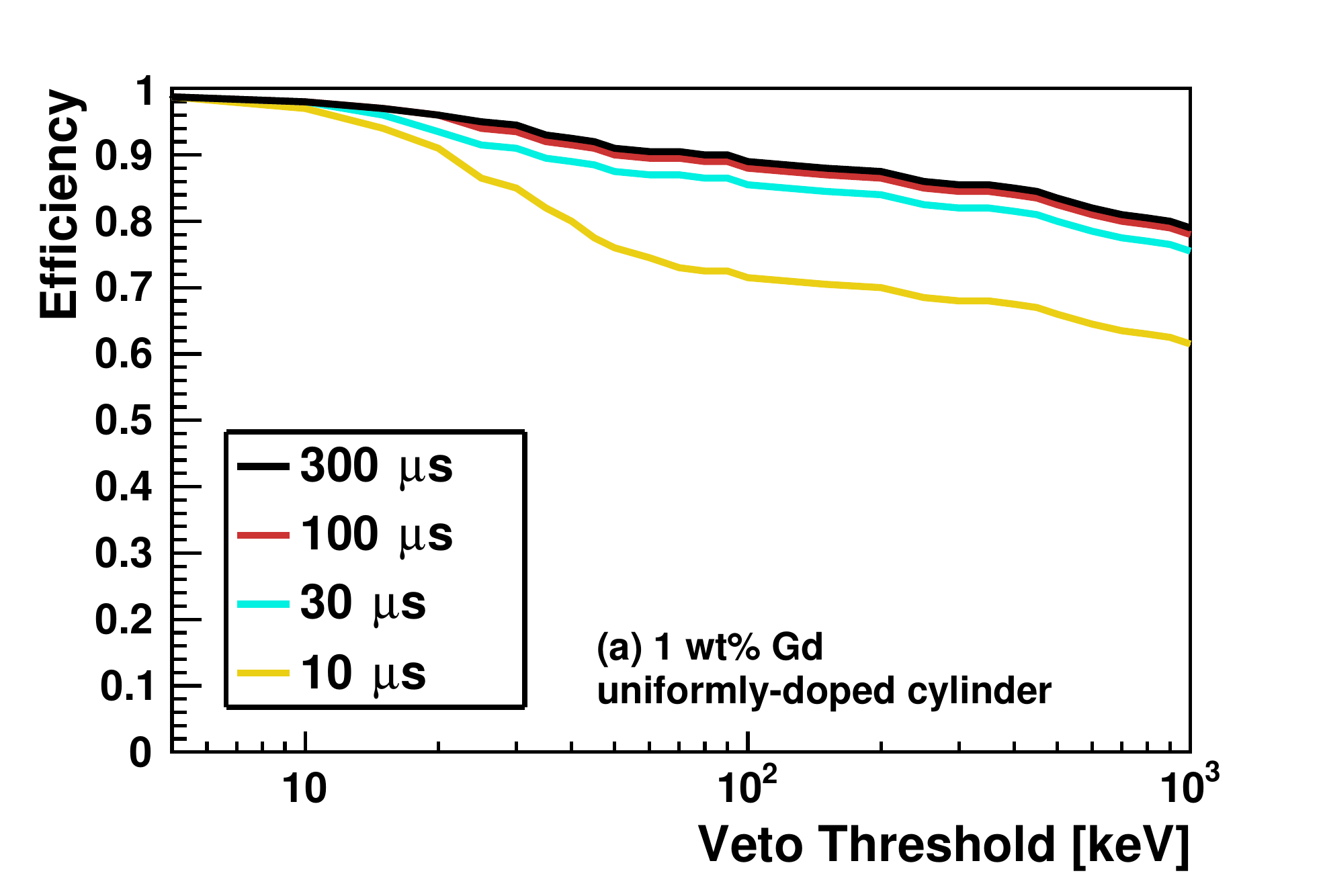}
\end{subfigure}
\begin{subfigure}{0.495\textwidth}  
	\centering
	\includegraphics[width = 0.95\columnwidth]{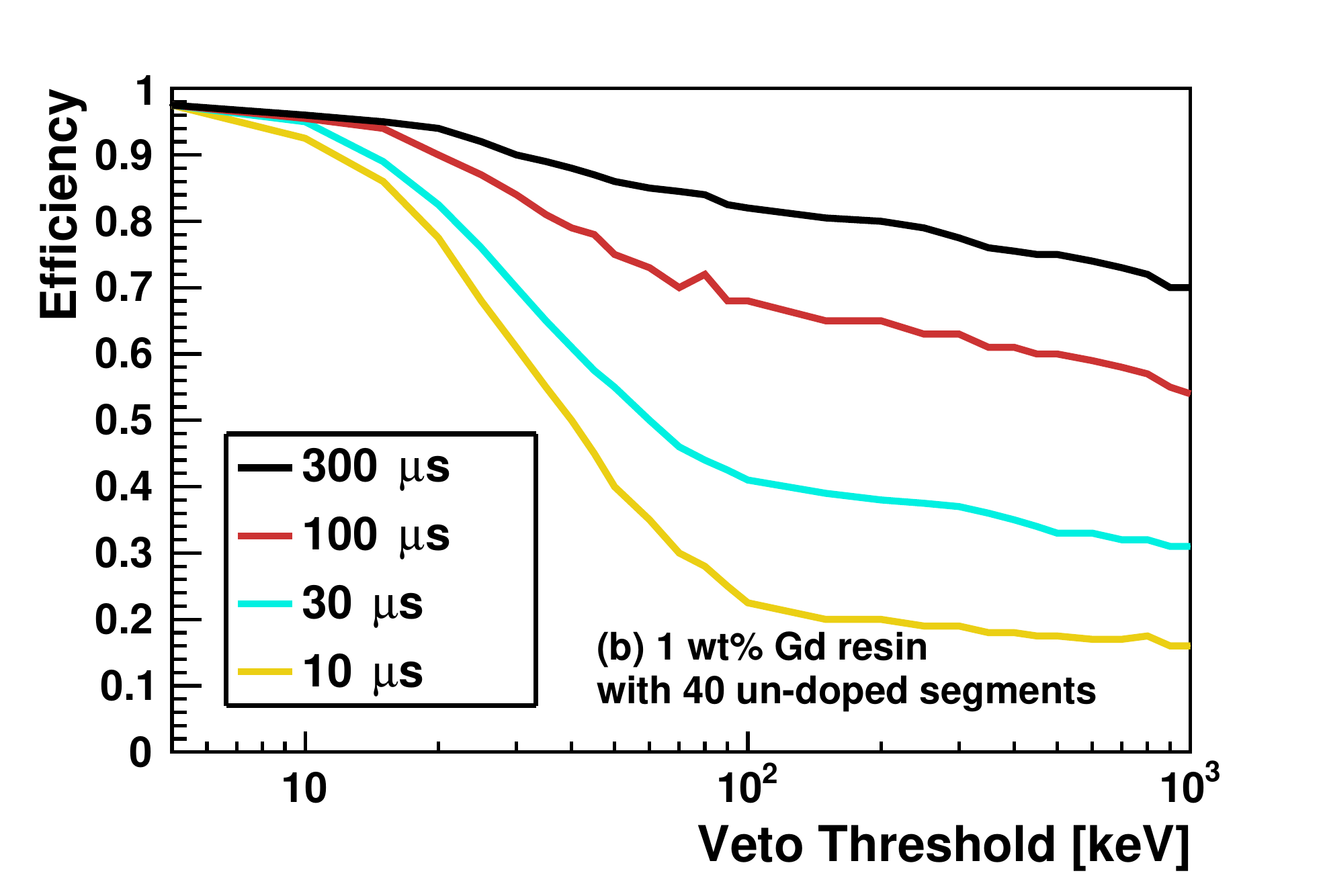}
\end{subfigure}
\begin{subfigure}{0.495\textwidth}  
	\centering
	\includegraphics[width = 0.95\columnwidth]{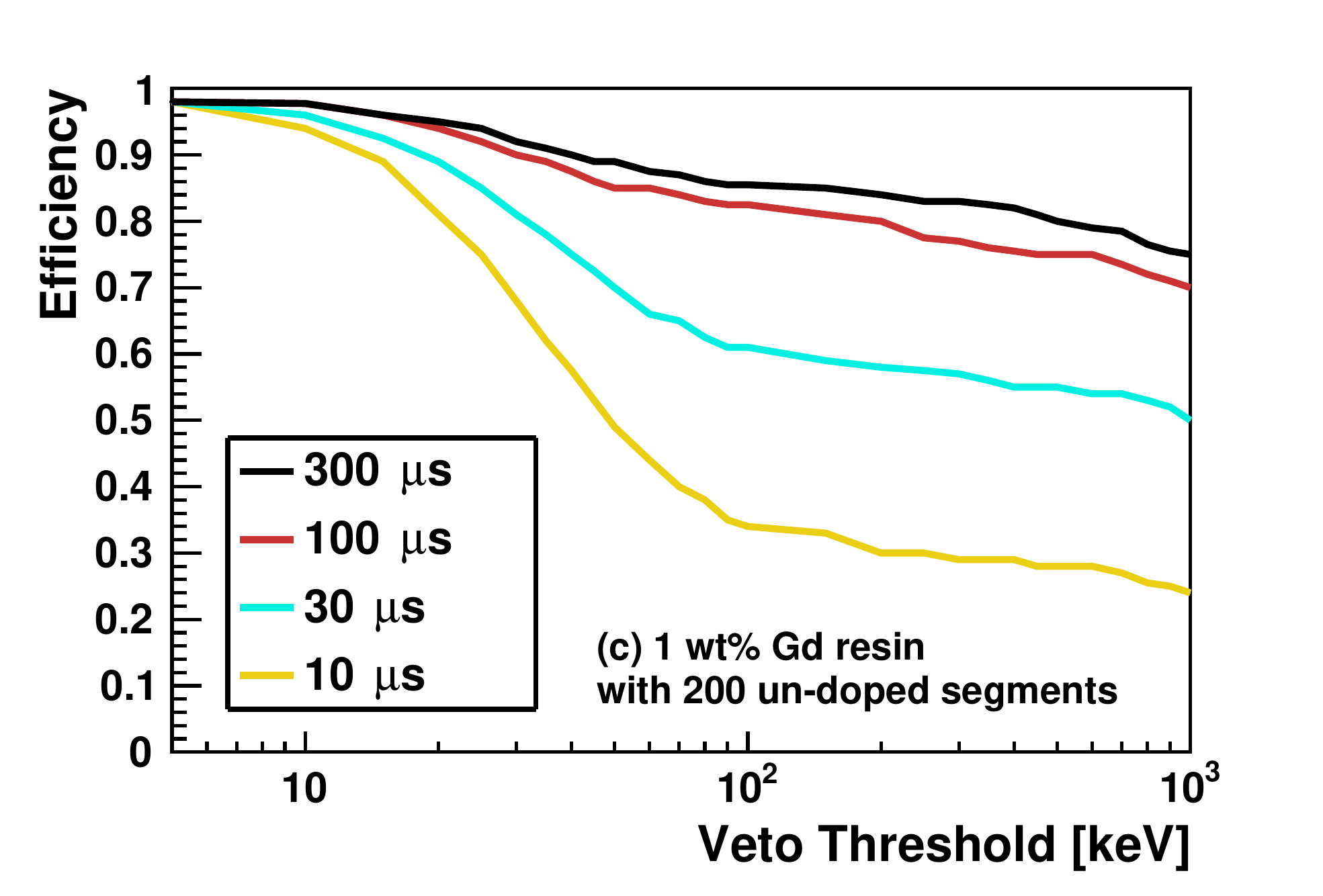}
\end{subfigure}
\begin{subfigure}{0.495\textwidth}  
	\centering
	\includegraphics[width = 0.95\columnwidth]{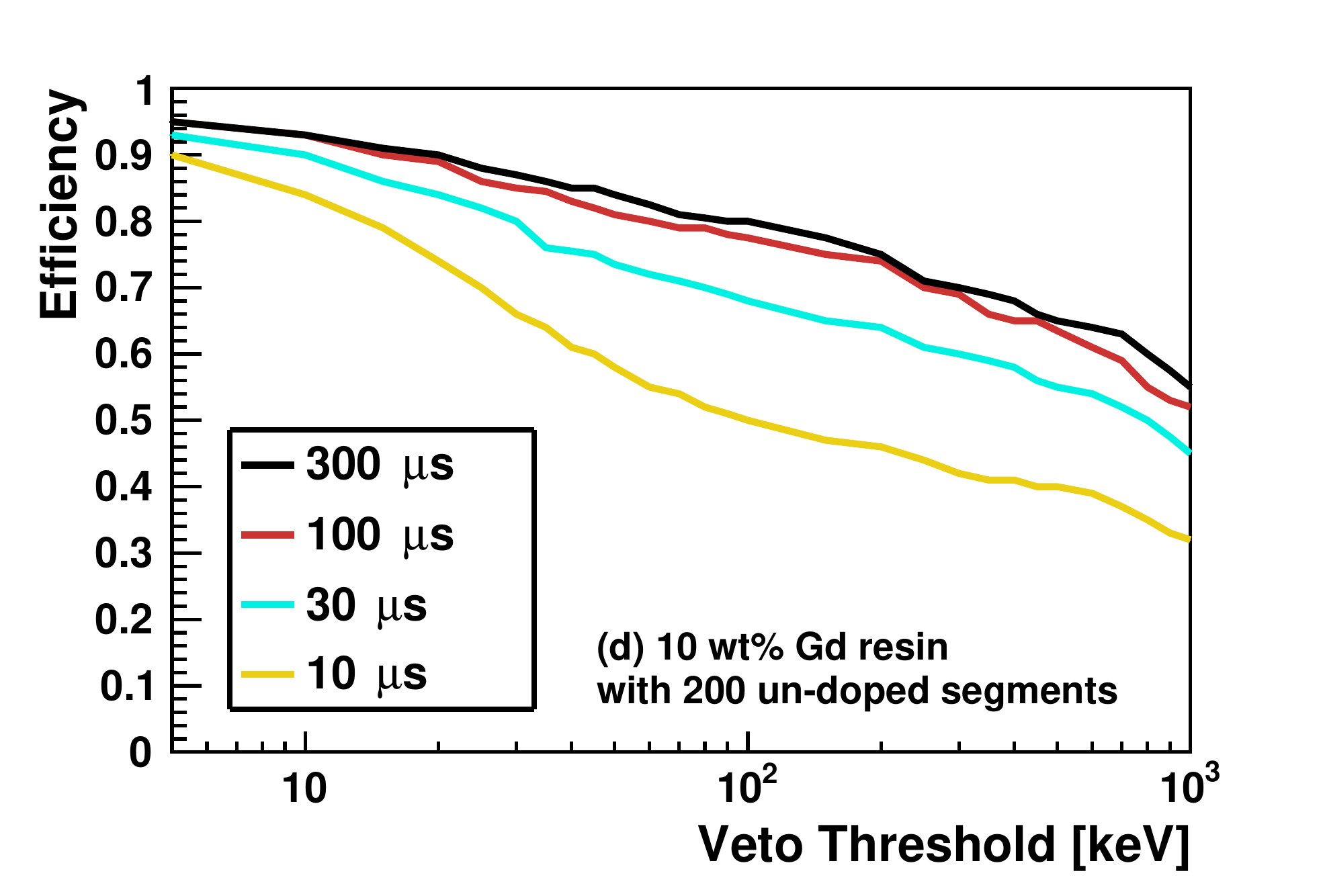}
\end{subfigure}
\caption{Neutron veto efficiencies for several configurations as a function of integration time and energy threshold. }
\label{fig:veto_efficiencies}
 \end{figure*}

Table \ref{tab:thick} gives the thicknesses of scintillator (outer arc length) and resin for a set of segmentation choices where the combination of scintillator and resin has an equivalent overall gadoliniujm mass of 1\% or 10\%, following the size constraints described above. These were compared to 1\% by mass uniform doping of the scintillator.  The veto efficiency curves are found in \FIG \ref{fig:veto_efficiencies}, each with four collection times, and plotted with respect to the veto threshold energy.  The lower the threshold energy and the longer the collection time,  the more likely a single-scatter nuclear recoil can be recognized above background.  

\begin{table}[t]
	\centering
	\setupthecaption
	\begin{tabular}{c|c|c|c}
	Mass Equiv. Gd & Segments & Scintillator & Gd Resin \\ \hline
	1\% & n=40 & 10.57 cm & 0.17 cm \\
	1\% & n=100 & 4.17 cm & 0.097 cm \\
	1\% & n=200 & 2.07 cm & 0.056 cm \\
	10\% & n=40 & 6.62 cm & 2.07 cm \\
	10\% & n=100 & 2.07 cm & 1.13 cm \\
	10\% & n=200 & 0.91 cm & 0.63 cm \\
	\end{tabular}
	\caption{Thicknesses of scintillator and resin for various segmentations with 1\% or 10\% mass equivalent Gd.}
	\label{tab:thick}
\end{table}

Note that the uniformly doped veto (\FIG \ref{fig:veto_efficiencies}(a)) has an efficiency of 90\%  with a reasonable choice of veto parameters such as a 100~$\mu$s timing window and a 50~keV threshold, but even for a very short 10~$\mu$s time window, efficiencies of 80\% are possible.   Looking at the segmented vetoes for 1\WT equivalent doped Gd (\FIGS \ref{fig:veto_efficiencies}(b) and \ref{fig:veto_efficiencies}(c)), the shortest timing windows never approach the efficiency of the uniformly doped configuration, even with increased segmentation.  However, for a collection time of 100~$\mu$s (and 50~keV threshold), a scintillator-to-resin ratio of 35:1 (calculated from Table \ref{tab:thick}) can come within 5\% of the veto efficiency of a uniformly-doped veto of mass equivalent 1\%.

\FIG \ref{fig:veto_efficiencies}(d) demonstrates what happens if the resin is made thicker (and the quantity of gadolinium goes up to 10\% mass equivalent).    For long collection times, the efficiencies lie below those for 1\WT Gd-loading.   The fact that the 10\WT Gd segmented veto efficiency never reaches 100\% is due to the decrease in the amount of scintillator and the fact that the resin absorbs the lower energy gammas, giving up some of the advantages of a low threshold.  Thus, if a short time window is necessary to tag single-scatter nuclear recoil events and the energy threshold is limited, the thicker resin can be an advantage.  There are also practical considerations. For 40 segments at the 10\WT Gd, the resin thickness becomes 2 cm  (scintillator:resin ratio of 3:2), making it impossible to consider most fabrication techniques (e.g. paint, foils, or laminates). 

Based on the simulations presented here for an active neutron veto of the type studied for SuperCDMS SNOLAB, it is possible to form an active veto from wedges of layered un-doped scintillator and Gd-loaded resin of mm thickness, with an efficiency approaching (within 5\%) that of a uniformly Gd-doped cylinder of the same Gd mass fraction.

\section{Conclusion and Future Efforts}

This wide-ranging study explores a variety of aspects related to producing solid gadolinium-loaded scintillators for neutron detection.  A \geant feasibility study using the SuperCDMS SNOLAB geometry showed that neutrons with a single scatter in the detector can be identified by their capture gammas in gadolinium-loaded scintillator with high efficiency on an event-by-event basis.  The simulated neutron veto was located at the site of the passive polyethylene shield to represent a possible upgrade path without disturbing the overall shield design.   

Small samples of doped scintillators were produced in our lab using a variety of commercially available compounds.  Practical details on how to make such samples are presented, along with the failure modes associated with certain compounds.    The most promising compounds are Gd(i-Pr)$_3$ and Gd(TMHD)$_3$, but  production in our lab demonstrated that the simple formula and process described in the literature~\CITE{Ovechkina,Bertrand} are not sufficient for achieving the concentrations required by our neutron veto simulation.   

The light output, attenuation length, and spectra were measured for our own samples and for the more highly-concentrated proprietary samples.  Addition of a small amount of gadolinium was observed to increase light output in some cases, but always decreases the attenuation length.  The spectrum can also shift to the green, depending on the specific compound.   Measurements on our small samples provided data which helped us to develop a very detailed optical photon \geant Monte Carlo.   The absolute number of detected photoelectrons is in reasonable agreement with the measurement, with the appropriate choice of variables such as reflectivity, absorption and emission spectra, QE, bulk attenuation, and fiber mask configuration.   The measured effective attenuation length of the small samples ($\lambda_{eff}$) provided the best leverage for tuning Monte Carlo variables.  A number of important variables, such as fiber trapping efficiency, can be determined once a tuned and reliable Monte Carlo is developed.   

In partnership with CEA Saclay, we plan to use the tools we have developed to find the optimum loading of gadolinium in plastic scintillator as a compromise between reduced light output and neutron capture.   The ability to cleanly cast larger scintillator elements blocks is under development.  Eventually the tools developed here will inform the construction of practical neutron vetos customized for the experiment in question. A practical solid scintillator alternative for neutron detection would provide a welcome alternative for wide range of rare-event searches.    

%\begin{acknowledgments}
\section*{Acknowledgements}
SuperCDMS is supported by a DOE Cosmic Frontier grant DE-SC0012294. This study was made possible by the Undergraduate Research Opportunities Program at the University of Minnesota and our excellent undergraduate students, three of whom are authors on this paper.  Acknowledgment and thanks also go to new students  T. Argawal and J. Nelson for help in fabricating samples and running the \geant simulation.  Professor Marc Hillmyer and Jake Brutman (Polymer Chemistry) were instrumental in early discussions and sample production setup.
%\end{acknowledgments}

%% BIBLIOGRAPHY
%\bibliographystyle{apsrev4-1}
\bibliographystyle{elsarticle-num} 
\bibliography{gd_bibliography}
%\input{neutron_detection.bbl}
%\bibliography{}

\end{document}